\documentclass[journal]{IEEEtran}

\usepackage[T1]{fontenc}
\usepackage{amsmath}
\usepackage{amssymb}
\usepackage{graphicx}
\usepackage{booktabs}
\usepackage{array}
\usepackage{cite}
\usepackage{orcidlink}

\graphicspath{{figures/}{./}{../../matlab/figures/}{../../figs/}{../figures/}{../}}

\usepackage{hyperref}
\hypersetup{
  hidelinks,
  breaklinks = true,
  pdfauthor  = {Shoumik Debnath, Anup Kumar Modak, Sudipta Saha},
  pdftitle   = {Physics-Informed Surrogate-Guided Phase-Change Trimming of
                Group-Velocity Dispersion in Thin-Film Lithium Niobate}
}

\newcommand{\supp}{supplementary material}
\newcommand{\SbS}{Sb$_2$S$_3$}
\newcommand{\SbSe}{Sb$_2$Se$_3$}
\newcommand{\LN}{LiNbO$_3$}

\newcommand{\pskm}{ps$^2$\,km$^{-1}$}
\newcommand{\pcskm}{ps$^3$\,km$^{-1}$}
\newcommand{\dbcm}{dB\,cm$^{-1}$}
\newcommand{\um}{$\mu$m}
\newcommand{\us}{$\mu$s}

\newcommand{\btwo}{\beta_2}
\newcommand{\bthree}{\beta_3}
\newcommand{\neff}{n_\mathrm{eff}}
\newcommand{\TLN}{T_\mathrm{LN}}
\newcommand{\Wc}{W_\mathrm{core}}
\newcommand{\Wp}{W_\mathrm{PCM}}
\newcommand{\Tp}{T_\mathrm{PCM}}

\begin{document}

\title{Physics-Informed Surrogate for Sb$_2$S$_3$ Dispersion Trimming in Thin-Film Lithium Niobate}

\author{Shoumik~Debnath$\orcidlink{0009-0001-9500-5319}^{1,2,\ast}$, Anup~Kumar~Modak$\orcidlink{0009-0003-8865-274X}^{2}$ and
  Sudipta~Saha\,\orcidlink{0009-0003-9110-8059}$^{1,3,\ast}$\\[6pt]
  {\small
  $^{1}$Department of Electrical and Electronic Engineering, Bangladesh
  University of Engineering and Technology, Dhaka 1205, Bangladesh\\
  $^{2}$Department of Computer Science and Engineering, Daffodil International
  University, Dhaka, Bangladesh\\
  $^{3}$Accident Research Institute, Bangladesh University of Engineering and
  Technology (BUET), Dhaka 1000, Bangladesh\\[4pt]
  $^{\ast}$Corresponding authors:
  \texttt{debnath.shoumik.buet@gmail.com} (S.~Debnath),
  \texttt{sudiptasaha@ari.buet.ac.bd} (S.~Saha)\\
  E-mail: \texttt{anupkumar.cse@diu.edu.bd} (A.~K.~Modak)}%
}

\markboth{}{}

\maketitle

\begin{abstract}
A film-thickness error of 8~nm, within the tolerance of commercial thin-film
lithium niobate wafers, shifts the group-velocity dispersion $\btwo$ at 1550~nm by
about 28~\pskm. We study numerically whether a segmented, non-volatile
phase-change cladding can correct this error after fabrication. In mode solves of
3012 geometries, \SbS{} switches $\btwo$ by a median of 181~\pskm{} at a 40~nm
strip thickness with no material absorption under the adopted optical constants,
whereas \SbSe{} reaches a comparable contrast at a median modal loss of
5.6~\dbcm. Because phase adds along the guide, four independently switched
binary segments reach a two-dimensional region of the $(\btwo,\bthree)$ plane;
eigenmode expansion supports this description at the junctions. Full-wave solves
at 65 die geometries show that the segment states move with the die error, so the
bit pattern is chosen from a per-die model: a spectral surrogate constrained by
exact identities of the lossless Maxwell problem. Its accuracy within the
training domain is similar to that of unconstrained baselines; it keeps $\btwo$
and $\bthree$ consistent, and its error at an unseen film thickness is 8.3 rather
than 13.0~\pskm. With a bare test ring and a second ring carrying a crystalline
strip already on the mask, 96.8\% of the dies are trimmed into $\pm25$~\pskm,
scored on full-wave solutions. A thermal model gives a candidate write window for
a 532~nm beam.
\end{abstract}

\vspace{0.5ex}

\section{Introduction}
\label{sec:intro}

\IEEEPARstart{T}{hin}-film \LN{} combines low propagation
loss~\cite{ZhangOptica2017} with a large second-order nonlinearity, a usable Kerr
coefficient and a fast electro-optic
response~\cite{WangNature2018,ZhuAOP2021,BoesScience2023}. Its nonlinear devices,
among them octave-spanning soliton
microcombs~\cite{SongLSA2024,GongOptica2020,HeOptica2019}, two-octave
supercontinuum~\cite{YuOL2019}, mid-infrared generation~\cite{MishraOE2022} and
broadband parametric amplification~\cite{JankowskiOptica2020}, are designed around
the group-velocity dispersion $\btwo$ and its slope $\bthree$. In a sub-micrometre
ridge the waveguide and material contributions to $\btwo$ are comparable and
opposite in sign, so these designs are specified to the nanometre.

A fabricated die rarely meets that specification. In the sweep reported below,
$\btwo$ at 1550~nm changes by 3.6~\pskm{} per nanometre of film thickness, so the
$\pm8$~nm thickness variation of a commercial wafer moves it by about 28~\pskm,
more than the $\pm25$~\pskm{} band used here as the target. Ridge width matters
about ten times less per nanometre, so the dominant error is set by the wafer
rather than by lithography. Fabrication-tolerant designs~\cite{HeftiAPLP2025}
lower the sensitivity at the cost of design freedom, but cannot correct a die
once it is made.

Post-fabrication trimming has been shown by localized
annealing~\cite{WuLPR2026,WuArXiv2025}, femtosecond-laser
writing~\cite{WangOL2022} and non-volatile phase-change
claddings~\cite{LianSUM2022,PangLPR2026}. The last hold their state without power,
and the low-loss chalcogenides \SbS{} and
\SbSe{}~\cite{DelaneyAFM2020,DelaneySciAdv2021} already support multi-bit
programmable circuits~\cite{ChenNatCommun2023,ChenIJEM2024} and foundry
integration~\cite{WeiNatCommun2024}. These schemes trim a scalar, a phase or an
effective index at one wavelength. Trimming dispersion is harder for two reasons.
The target has two components, $\btwo$ and $\bthree$, which one continuous control
cannot set independently. The trimming elements also sit on the same mis-sized
ridge, so their own effect shifts with the error they are meant to correct. We are
not aware of a report in which group-velocity dispersion is corrected after
fabrication.

\begin{figure*}[!t]
  \centering
  \includegraphics[width=\textwidth]{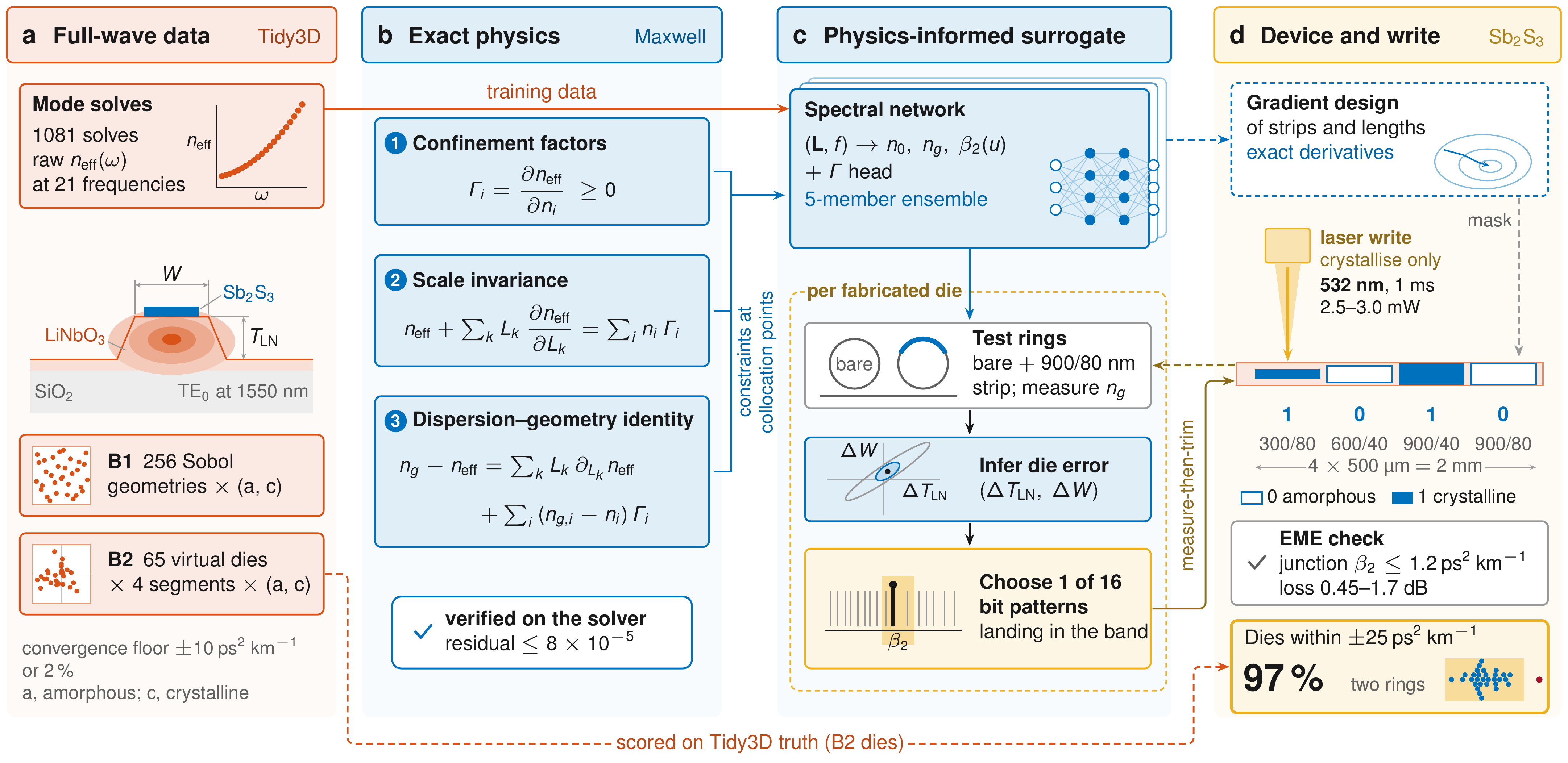}
  \caption{Workflow of the study. (a)~Full-wave data: 1081 mode solves with raw
    $\neff(\omega)$, from the space-filling (B1) and jitter (B2) sets of
    Sec.~\ref{sec:p2data}. (b)~Exact identities of Maxwell's equations imposed on
    the surrogate (Sec.~\ref{sec:surrogate}). (c)~The physics-informed surrogate
    and the per-die loop: test rings, inferred die error and the choice of 1 of 16
    bit patterns, with gradient design of the segment library as a side branch.
    (d)~The eigenmode-expansion check of the junctions, the success rate scored
    on the B2 dies, and the thermal estimate of the laser write.}
  \label{fig:workflow}
\end{figure*}

\begin{figure*}[htbp]
  \centering
  \includegraphics[width=\textwidth]{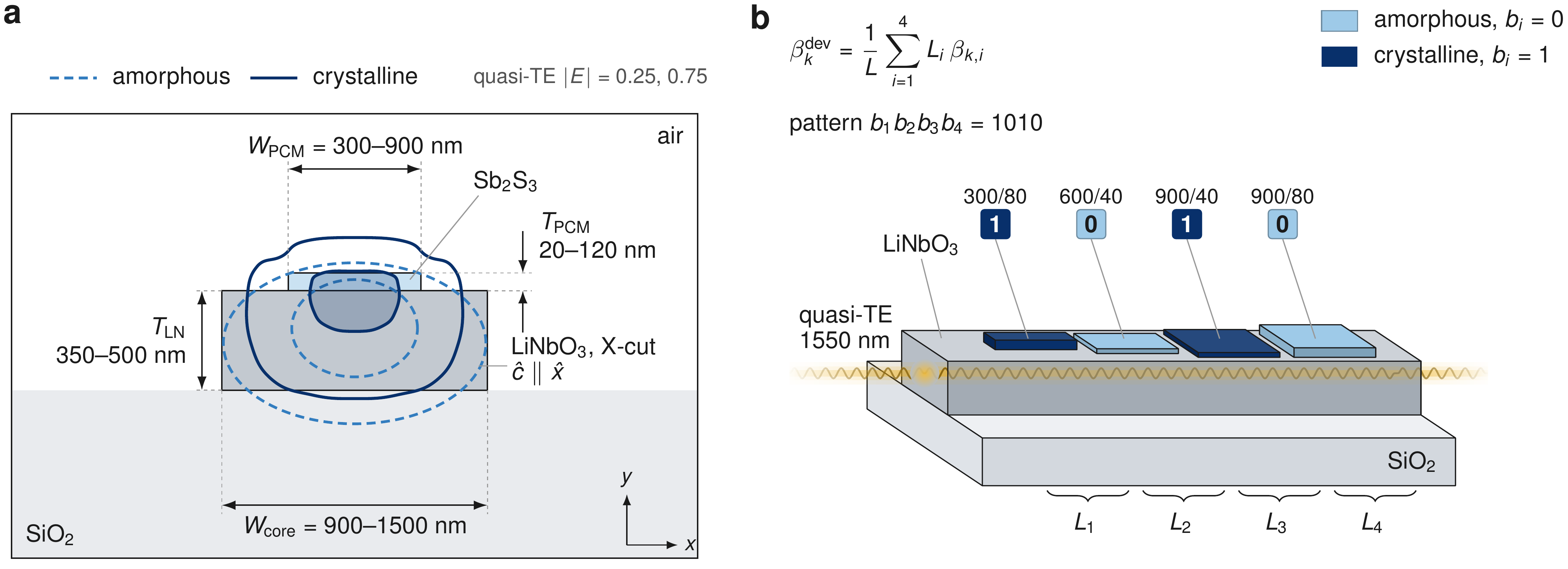}
  \caption{Waveguide and trimmer. (a)~Cross-section of the X-cut \LN{} ridge with
    the \SbS{} strip, to scale at $\TLN=450$, $\Wc=1200$, $\Wp=600$ and
    $\Tp=80$~nm, with the swept dimensions labelled. Contours mark $|E|=0.25$
    and 0.75 of the quasi-TE mode at 1550~nm, strip amorphous (dashed) and
    crystalline (solid). (b)~The four-segment trimmer of Sec.~\ref{sec:yield}
    ($\Wp/\Tp=300/80$, 600/40, 900/40 and 900/80~nm) in bit pattern 1010
    ($1$: crystalline). Segment lengths are not to scale.}
  \label{fig:geom}
\end{figure*}

Here we study, by simulation, post-fabrication trimming of dispersion in an X-cut
thin-film \LN{} ridge with a segmented \SbS{} cladding
(Figs.~\ref{fig:workflow} and~\ref{fig:geom}), and address both difficulties.
Because phase adds along the guide, independently switched binary segments give
two-dimensional control of $(\btwo,\bthree)$. Mode solves at 65 die geometries
show that the segment states do not shift rigidly with the die error, which
limits a rule based on the nominal library to 83\% of these dies even when the
error is known exactly. We therefore choose the bit pattern from a per-die model,
a spectral surrogate constrained by exact identities of the lossless Maxwell
problem, and infer the die error from two test rings. One of them carries a
crystalline strip already on the mask, which removes most of the near-degeneracy
between film and width errors. Scored on full-wave solutions, this procedure
brings 96.8\% of the dies within $\pm25$~\pskm. We also report its sensitivity to
a systematic solver offset and a thermal estimate of the laser write; numerical
details are in the \supp{}.

\section{Device, Material Models and Numerical Methods}
\label{sec:device}

\subsection{Waveguide and materials}

The waveguide is a fully etched X-cut \LN{} ridge on silica, with a phase-change
strip on the top surface and air above (Fig.~\ref{fig:geom}). The optic axis lies
in the film plane, so the dominant field component of the quasi-TE fundamental
mode lies along the extraordinary axis. Four dimensions are swept: the film
thickness $\TLN$ (350 to 500~nm in 50~nm steps), the ridge width $\Wc$ (900 to
1500~nm in 100~nm steps), and the strip width $\Wp$ (300, 600 and 900~nm) and
thickness $\Tp$. The strip is no wider than the ridge, so the mode is guided by
the ridge and perturbed by the strip. Because the two materials differ in index
contrast, $\Tp$ is 20, 40, 80 and 120~nm for \SbS{} and 10, 20, 40 and 60~nm for
\SbSe{}. A thicker strip gives a larger switching contrast but pulls the mode out
of the \LN{} and, for \SbSe{}, raises the loss.

\SbS{} and \SbSe{} switch reversibly between amorphous and crystalline phases
with a large index contrast and, unlike the tellurides, without a strong
absorption penalty in the telecom band~\cite{DelaneyAFM2020}. The \LN{} film uses
the extraordinary-index Sellmeier fit of Zelmon
\textit{et al.}~\cite{ZelmonJOSAB1997}, and the phase-change layers use the
ellipsometry $n$ and $k$ of Delaney \textit{et al.}~\cite{DelaneyAFM2020} for the
two end states (Fig.~S1 in the \supp{}). At 1550~nm the \SbS{} index rises from
2.72 to 3.31 on crystallisation and the \SbSe{} index from 3.29 to 4.05. For
\SbS{}, $k$ is zero throughout the band in both states; crystalline \SbSe{} has
$k$ from $5.2\times10^{-4}$ at 1400~nm to $2.1\times10^{-4}$ at 1550~nm, the
origin of the \SbSe{} losses reported here. Measured \SbS{} constants vary
between laboratories with deposition conditions and
stoichiometry~\cite{GutierrezIScience2022}; another source would shift the
absolute dispersion values rather than the trends.

Intermediate crystalline fractions $f$ follow the Lorentz--Lorenz
effective-medium rule,
\begin{equation}
  \frac{\varepsilon_\mathrm{eff}-1}{\varepsilon_\mathrm{eff}+2}
  = f\,\frac{\varepsilon_\mathrm{c}-1}{\varepsilon_\mathrm{c}+2}
  + (1-f)\,\frac{\varepsilon_\mathrm{a}-1}{\varepsilon_\mathrm{a}+2},
  \label{eq:lorentzlorenz}
\end{equation}
applied to the complex permittivities $\varepsilon_\mathrm{a}$ and
$\varepsilon_\mathrm{c}$ of the end states at $f=0$, 0.25, 0.5, 0.75 and 1. The
device of Sec.~\ref{sec:trim} uses only the end states; the intermediate points
show how non-linear the response is between them.

\subsection{Mode solver, dispersion extraction and the numerical floor}
\label{sec:conv}

Modes were computed with the finite-difference mode solver in Tidy3D
2.11.2~\cite{Tidy3D} on a 6.0 $\times$ 4.0~\um{} domain at 32 grid steps per
wavelength, with every layer boundary on a cell edge (Fig.~S2). Each geometry was
solved at 11 frequencies spaced uniformly in $\omega$ across 1380 to 1720~nm, the
1400 to 1700~nm reporting band extended by 20~nm at each end. A polynomial in
$\omega$ was fitted to $\neff(\omega)$ and differentiated analytically to give
$\btwo=\mathrm{d}^2\beta/\mathrm{d}\omega^2$ and
$\bthree=\mathrm{d}^3\beta/\mathrm{d}\omega^3$, with the order raised only while
the fit residual continued to halve.

Each dispersive medium is fitted to a pole-residue model before the solve, and a
small RMS error in the index can be a large error in its curvature: in a
diagnostic run with a tolerance of $2\times10^{-4}$, the bulk \LN{} $\btwo$ at
1550~nm came out at 373~\pskm, against 99.5~\pskm{} at $10^{-7}$ and 99.2~\pskm{}
from a spline through the raw data (Table~S1). Every production solve uses the
$10^{-7}$ tolerance.

Convergence was assessed at a geometry on a zero crossing, where only an absolute
criterion is meaningful (Table~S2). Shrinking the domain or lowering the
resolution moves $\btwo$ by 41.9 and 56.1~\pskm, whereas enlarging the domain moves
it by 0.5~\pskm, so the settings used are the smallest that are converged. A
re-check on the second dataset against a finer reference keeps every amorphous
variation within 4~\pskm{} and brings the crystalline 900/80~nm strip within
1~\pskm{} once the strip is resolved by twice as many cells (Sec.~S2). We take the
numerical floor as the larger of $\pm10$~\pskm{} and 2\% of $|\btwo|$, and quote no
zero-dispersion wavelength to better than $\pm10$~nm. As a property of the
discretisation, this residual is expected to be largely systematic, correlated
between neighbouring geometries on the same mesh rather than random from die to
die; Sec.~\ref{sec:offset} treats it as a common offset on every die.

\subsection{Dataset and quality filter}

A solved geometry enters the dataset only if a guided quasi-TE mode exists, if
$\neff$ falls monotonically with wavelength across the extended band, which
rejects anticrossings with a strip-guided mode, and if the $\neff$ fit residual is
at most $10^{-6}$. The sweep produced 3012 converged solutions, 1546 for \SbS{}
and 1466 for \SbSe{}, covering 92.0\% and 87.3\% of their grids (Table~S3). The
rejections concentrate at the thickest strip, and no reported result rests on an
excluded node.

\subsection{A second, full-wave dataset for the surrogate and the die-level checks}
\label{sec:p2data}

The sweep stores $\btwo$ but not the raw $\neff(\omega)$, and it samples a fixed
nominal ridge, so it cannot show how the segment states move when the ridge is
mis-sized. A second set of solves with the same solver, materials and quality
filter keeps the raw $\neff$, $k_\mathrm{eff}$ and TE fraction at 21 frequencies
across 1380 to 1720~nm (Fig.~S3). Its space-filling part draws 256 geometries
from a scrambled Sobol sequence~\cite{Sobol1967} over $\TLN=400$ to 500~nm,
$\Wc=1000$ to 1400~nm, $\Wp=300$ to 900~nm ($\Wp\le\Wc$) and $\Tp=20$ to 120~nm,
each solved with the strip amorphous and crystalline. Its jitter part re-solves
the four mask segments of Sec.~\ref{sec:yield} in both states, and the bare ridge,
at 65 die geometries: the nominal ridge and 64 Sobol points in the box
$|\Delta\TLN|\le3\sigma_T$, $|\Delta\Wc|\le3\sigma_W$ with $\sigma_T=8$~nm and
$\sigma_W=14.9$~nm. Every value used in Sec.~\ref{sec:dietruth} is a mode
solve at the die's own geometry. Of the 1097 solves, 1081 passed the filter: all
585 jitter solves and all but 16 of the 512 space-filling solves.

Every solve also stores, at each frequency, the confinement factors of
Eq.~\eqref{eq:gamma} for the extraordinary and ordinary components in the \LN{},
the strip, the silica and the air. At 1550~nm, $\Gamma_{\mathrm{LN},e}$ ranges
from 0.45 to 0.92, $\Gamma_{\mathrm{LN},o}$ from 0.06 to 0.13 and
$\Gamma_\mathrm{PCM}$ up to 0.53. The group index rebuilt from them as the energy
velocity, $\sum_i n_{g,i}\Gamma_i$, matches the fitted group index to a median of
$2.8\times10^{-3}$.

\section{Dispersion Tuning and Material Selection}
\label{sec:tuning}

\subsection{Dispersion of the as-deposited waveguide}
\label{sec:asfab}

With the strip amorphous, $\btwo$ at 1550~nm is normal for thin films and
anomalous for thick ones. Over the \SbS{} sweep the median value falls from
$+419$~\pskm{} at $\TLN=350$~nm to $+130$ at 400~nm, $-63$ at 450~nm and
$-202$~\pskm{} at 500~nm. The zero crossing lies between 400 and 450~nm for most
widths, and its locus runs diagonally across the $(\Wc,\TLN)$ plane (Fig.~S4): a
thicker film can be brought back to zero dispersion by a narrower ridge.

Between $\TLN=400$ and 450~nm the median slope is $-3.6$~\pskm{} per nanometre
(interquartile range $-4.0$ to $-3.1$), and the median slope in ridge width is
0.56~\pskm{} per nanometre. An 8~nm thickness error therefore moves $\btwo$ by
about 28~\pskm{} and a 15~nm width error by about 8~\pskm.

Widening the strip from 300 to 900~nm moves the zero locus toward thinner
phase-change layers (Fig.~S5). Counting every crystalline fraction, a
zero-dispersion wavelength inside 1400 to 1700~nm occurs for 16.8\% of \SbS{}
solutions and 18.9\% of \SbSe{} solutions (21.9\% and 28.2\% as deposited).
Figure~S6 gives the distributions of $\neff$, $\btwo$, mode area and loss.

\subsection{Switching contrast, loss and the choice of material}
\label{sec:material}

\begin{figure*}[!t]
  \centering
  \includegraphics[width=0.8\textwidth]{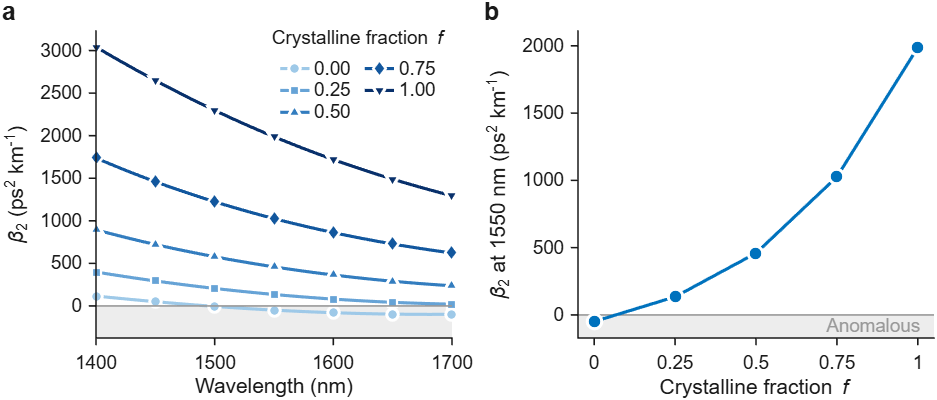}
  \caption{Dispersion switching of one \SbS{}-clad waveguide ($\TLN=450$,
    $\Wc=1200$, $\Wp=600$, $\Tp=80$~nm). (a)~$\btwo(\lambda)$ at five crystalline
    fractions. (b)~$\btwo$ at 1550~nm against $f$. Shaded: anomalous
    dispersion.}
  \label{fig:gvd}
\end{figure*}

Crystallising the strip raises the cladding index and pulls the mode upward,
changing both $\neff$ and its curvature. For a representative \SbS{} geometry
($\TLN=450$, $\Wc=1200$, $\Wp=600$, $\Tp=80$~nm), $\btwo$ at 1550~nm moves from
$-49$~\pskm{} amorphous to $+1988$~\pskm{} fully crystalline, passing 135, 459 and
1026~\pskm{} at $f=0.25$, 0.5 and 0.75 (Fig.~\ref{fig:gvd}). Because the response
is strongly non-linear in $f$, the device uses only the two end states.

\begin{figure*}[!t]
  \centering
  \includegraphics[width=0.9\textwidth]{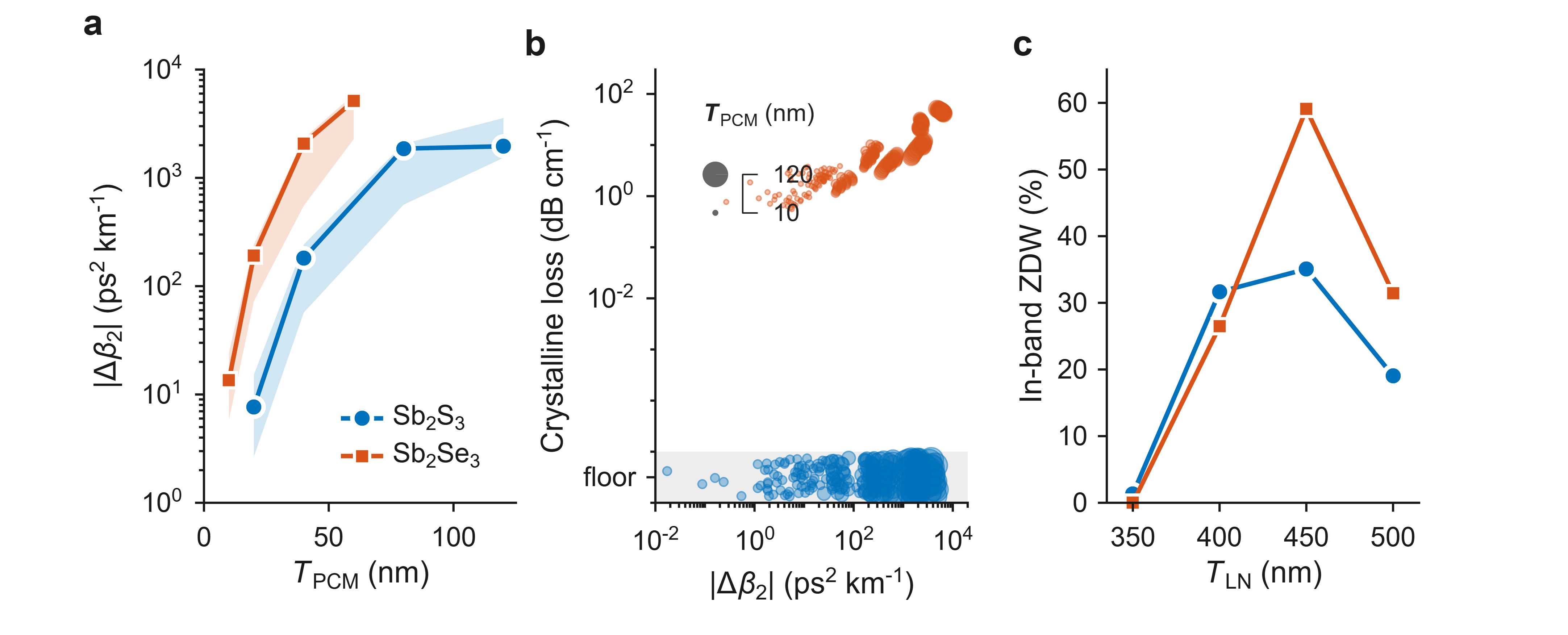}
  \caption{Switching range and its cost. (a)~Median $|\Delta\btwo|$ against strip
    thickness; bands: interquartile range. (b)~Switching contrast against
    crystalline-state loss, bubble area $\propto\Tp$; \SbS{} losses ($k=0$ in the
    adopted data) sit at the numerical floor (shaded). (c)~Fraction of as-deposited geometries with an
    in-band zero-dispersion wavelength, against film thickness.}
  \label{fig:matcompare}
\end{figure*}

Switching contrast scales sharply with strip thickness
[Fig.~\ref{fig:matcompare}(a)]. The median $|\Delta\btwo|$ for \SbS{} rises from
7.6~\pskm{} at $\Tp=20$~nm to 181 at 40~nm, 1863 at 80~nm and 1957~\pskm{} at
120~nm; \SbSe{} reaches similar values at roughly half the thickness, consistent
with its larger index contrast.

The materials separate on loss [Fig.~\ref{fig:matcompare}(b)]. With the
extinction coefficient of the adopted dataset, zero for \SbS{} in both
states~\cite{DelaneyAFM2020}, the computed modal loss of the \SbS{} sweep never
exceeds $2.7\times10^{-4}$~\dbcm, below the level at which the pole fit is
meaningful; crystalline \SbSe{} has a median of 5.6~\dbcm{} and a maximum of
51~\dbcm. The \SbS{} value reflects the optical constants used, not a lossless
film: residual absorption of deposited films and edge scattering are not
modelled, and junction loss is treated in Sec.~\ref{sec:eme}. Within these
constants \SbS{} provides the range within a loss budget compatible with a
resonator or a long nonlinear interaction (Table~\ref{tab:matcompare}), and the
rest of the paper uses it.

\begin{table}[!t]
  \centering
  \caption{The two claddings over the converged sweep and the linear Monte Carlo
    of Sec.~\ref{sec:yield} (four-segment set on the mask). Losses are modal
    propagation losses computed with the optical constants of
    Ref.~\cite{DelaneyAFM2020} ($k=0$ for \SbS{}).}
  \label{tab:matcompare}
  \begin{tabular}{@{}p{0.46\columnwidth}cc@{}}
    \toprule
     & \SbS{} & \SbSe{}\\
    \midrule
    Index at 1550~nm, amorphous to crystalline & 2.72 to 3.31 & 3.29 to 4.05\\
    Strip thicknesses swept (nm) & 20, 40, 80, 120 & 10, 20, 40, 60\\
    Converged geometries (grid coverage) & 1546 (92.0\%) & 1466 (87.3\%)\\
    Solutions with an in-band zero crossing, all $f$ (as deposited) & 16.8\% (21.9\%) & 18.9\% (28.2\%)\\
    Median crystalline-state loss (\dbcm) & below floor & 5.6\\
    Largest computed loss, any state (\dbcm) & $2.7\times10^{-4}$ & 51\\
    Dies trimmed inside $\pm25$~\pskm, linear model & 99.66\% & 93.3\%\\
    Median added loss after trimming (\dbcm) & $1.2\times10^{-7}$ & 3.1\\
    Bit patterns used, of 16 & 7 & 16\\
    \bottomrule
  \end{tabular}
\end{table}

The sign of the swing depends on the strip thickness. At $\Tp\ge40$~nm,
crystallising \SbS{} moves $\btwo$ in the normal direction, so a device biased
anomalous can be trimmed toward zero. At $\Tp=20$~nm the mean swing over the
sensitivity set is $-25$~\pskm, in the anomalous direction, possibly because the
thin crystalline layer changes the curvature of the mode more than its
confinement. The segment library stays in the thick-strip regime, where the sign
is consistent.

\section{Segmented Binary Control}
\label{sec:trim}

\subsection{Dispersion of a segmented guide}

Consider a waveguide of total length $L$ divided into $M$ segments of length
$L_i$, each with its own independently switched strip. If the junctions are
adiabatic (checked in Sec.~\ref{sec:eme}), the accumulated phase is
\begin{equation}
  \phi(\omega) = \sum_{i=1}^{M} L_i\,\beta_i(\omega),
  \label{eq:phasesum}
\end{equation}
where $\beta_i$ is the propagation constant of segment $i$ in its current state.
Every $\omega$-derivative inherits the linearity, so the device dispersion of
order $k$ is
\begin{equation}
  \beta_k^{\mathrm{dev}}
  = \frac{1}{L}\sum_{i=1}^{M} L_i\,\beta_{k,i},
  \qquad L = \sum_i L_i .
  \label{eq:betamix}
\end{equation}
The device therefore has $2^M$ discrete dispersion states, each addressed by an
$M$-bit pattern. A single strip switched continuously in $f$ traces a curve in
the $(\btwo,\bthree)$ plane, so fixing $\btwo$ fixes $\bthree$; segments whose
switching vectors point in different directions span a region with area.

\subsection{The segment library and the reachable set}

A library of eleven \SbS{} segment types was solved on the nominal waveguide
($\TLN=450$~nm, $\Wc=1200$~nm), covering three strip widths and four strip
thicknesses in both states. Each entry is a $(\btwo,\bthree,\neff,\alpha)$ tuple
per state, and any candidate device follows from Eq.~\eqref{eq:betamix} without
further mode solving.

\begin{figure*}[!t]
  \centering
  \includegraphics[width=0.9\textwidth]{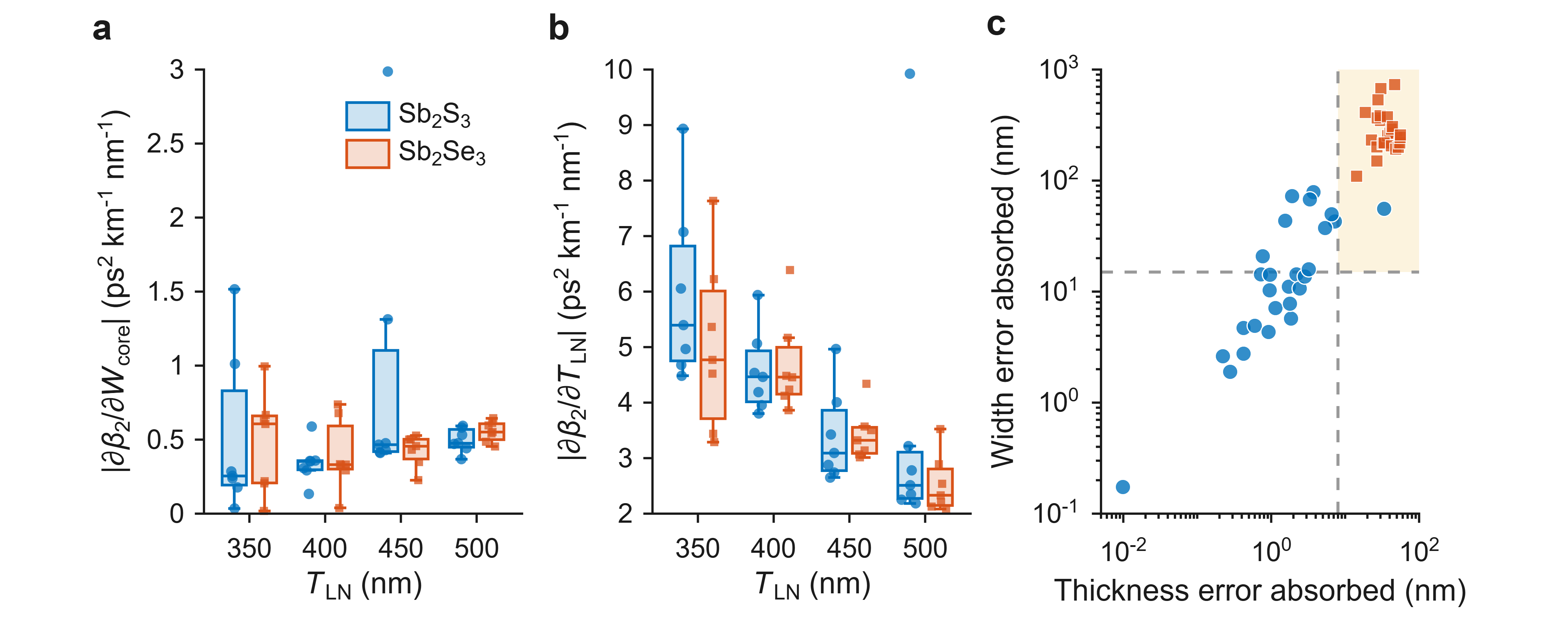}
  \caption{Fabrication sensitivity and trim authority. (a)~$|\partial\btwo/\partial\Wc|$
    and (b)~$|\partial\btwo/\partial\TLN|$ against film thickness, per material.
    (c)~Nanometres of each error that one segment absorbs; dashed lines are the
    $1\sigma$ process windows (15~nm in width, 8~nm in thickness).}
  \label{fig:sens}
\end{figure*}

Figure~\ref{fig:sens} compares the process sensitivities with the authority of
one segment, over 28 nominal geometries at the strip slice $\Wp=900$~nm,
$\Tp=20$~nm. The derivatives $\partial\btwo/\partial\Wc$ and
$\partial\btwo/\partial\TLN$ are properties of the ridge and vary slowly with the
strip, so this slice sets the error budget.

\begin{figure*}[!t]
  \centering
  \includegraphics[width=0.85\textwidth]{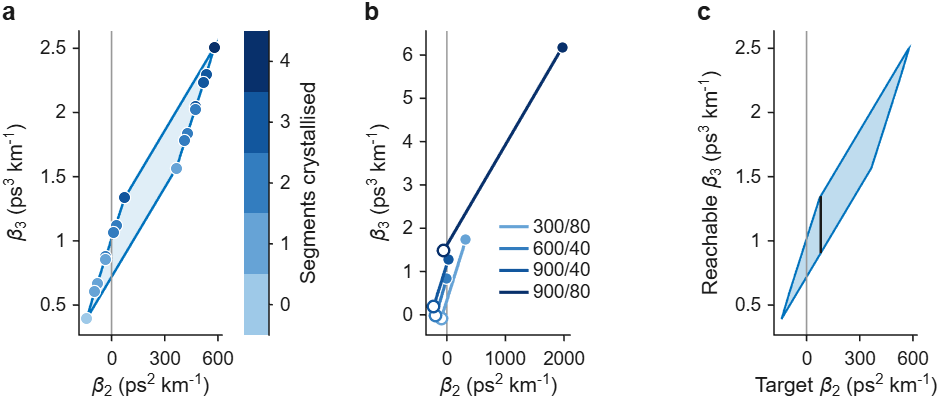}
  \caption{(a)~The 16 states of the four-segment trimmer in the
    $(\btwo,\bthree)$ plane, coloured by the number of crystallised segments,
    with their convex hull. (b)~Per-segment switching vectors from amorphous
    (open) to crystalline (filled). (c)~$\bthree$ interval reachable at each
    $\btwo$; the bar marks the widest, 0.44~\pcskm{} at $\btwo\approx80$~\pskm.}
  \label{fig:reach}
\end{figure*}

Enumerating the 16 bit patterns of the four segments selected in
Sec.~\ref{sec:yield} gives the point cloud of Fig.~\ref{fig:reach}(a). The set
spans $-140$ to 580~\pskm{} in $\btwo$ and 0.39 to 2.51~\pcskm{} in $\bthree$, and
the widest $\bthree$ interval at a single $\btwo$ is 0.44~\pcskm. The area comes
from segments moving the device in different directions through the plane
[Fig.~\ref{fig:reach}(b)].

\subsection{A full-wave check of the segment junctions}
\label{sec:eme}

Equations~\eqref{eq:phasesum} and~\eqref{eq:betamix} neglect reflection, mode
conversion and radiation where the strip changes. This was tested with eigenmode
expansion~\cite{Bienstman2001} in Tidy3D~\cite{Tidy3D}, with abrupt junctions and
each cross-section described by its three lowest modes, sorted by TE fraction and
solved at 11 frequencies between 1380 and 1720~nm; the fundamental mode of every
cross-section matches the mode solves to six decimal places in $\neff$. Each
segment is 100~\um{} long between 20~\um{} bare leads. The part of the
transmitted phase not explained by the length-weighted model was fitted with a
quartic in frequency, and its curvature is reported as a contribution to the
$\btwo$ of a device of four 500~\um{} segments.

\begin{figure*}[!t]
  \centering
  \includegraphics[width=0.9\textwidth]{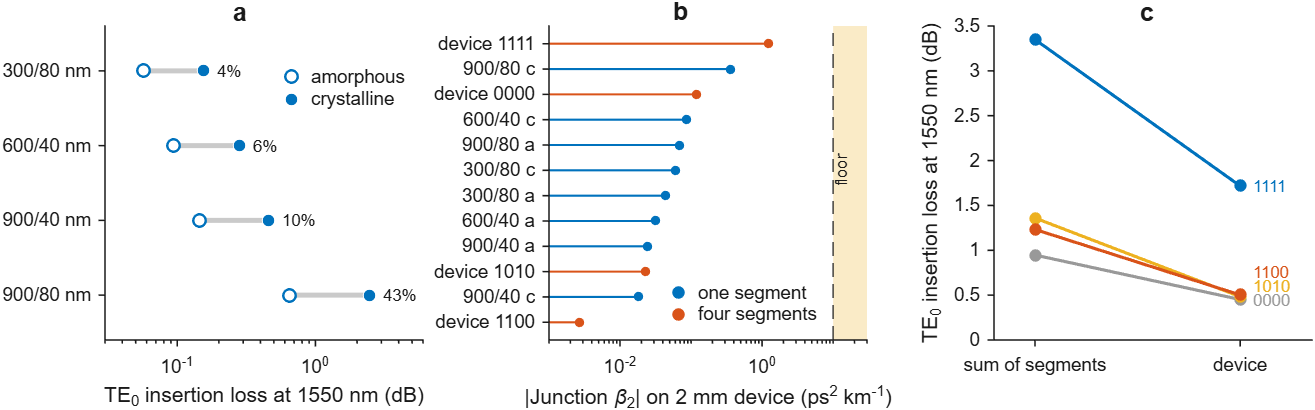}
  \caption{Eigenmode-expansion check of the segment junctions at 1550~nm.
    (a)~TE$_0$ insertion loss of each segment alone, amorphous (open) and
    crystalline (filled). (b)~Junction contribution to the $\btwo$ of the 2~mm
    device against the 10~\pskm{} floor. (c)~Loss of the four-segment device
    against the sum of its segments, per bit pattern.}
  \label{fig:eme}
\end{figure*}

The junctions add at most 0.36~\pskm{} to $\btwo$ for any single segment and
1.24~\pskm{} for the four-segment device in pattern 1111; for patterns 0000, 1010
and 1100 the contribution is 0.12, 0.02 and 0.003~\pskm{}
[Fig.~\ref{fig:eme}(b)]. All are an order of magnitude below the numerical floor,
so within that floor Eq.~\eqref{eq:betamix} holds for the 2~mm device.

A single segment between bare leads transmits TE$_0$ with a loss of 0.06 to
0.65~dB amorphous and 0.16 to 2.45~dB crystalline [Fig.~\ref{fig:eme}(a)]. The
largest loss belongs to the crystalline 900/80~nm strip, whose fundamental index,
2.03, differs most from that of the bare ridge, 1.79. Reflection stays below
$-21.5$~dB in every run, so the lost power is coupled to higher modes or radiated.
Inside the device the index steps between neighbouring strips are smaller, and
the four segments together lose 0.45~dB (0000) to 1.72~dB (1111), a third to a
half of the sum for isolated segments [Fig.~\ref{fig:eme}(c)]. With three modes
per cross-section these values indicate the scale of the junction loss rather
than converged figures. Tapers at the strongest segment, or the loss-weighted
library of Sec.~\ref{sec:libdesign}, would reduce it without changing
Eq.~\eqref{eq:betamix}.

\section{A Physics-Informed Surrogate for the Design Space}
\label{sec:surrogate}


The segment states do not move rigidly with the die error
(Sec.~\ref{sec:dietruth}), so choosing a pattern for a measured die needs the
library at the die's own geometry, and inferring the die's
$(\Delta\TLN,\Delta\Wc)$ from a test-ring group index needs
$\partial n_g/\partial\TLN$ and $\partial n_g/\partial\Wc$. A regression that fits
values well can still have derivatives of the wrong sign. The surrogate below
supplies both, with derivatives constrained by relations that hold exactly for
the Maxwell problem solved here.

\subsection{Exact relations for this waveguide}

The relations assume a non-magnetic, $z$-invariant guide whose media have
diagonal permittivity tensors aligned with the guide axes; those built on
Eq.~\eqref{eq:pert} also assume a lossless guide. Both hold exactly for the
training data: the \LN{} Sellmeier model, the constant-index silica and the
adopted \SbS{} constants ($k=0$) are lossless. The scale invariance used in
Eq.~\eqref{eq:sum1} holds with loss as well, but its expression through the
confinement factors does not, so a lossy cladding such as crystalline \SbSe{}
would need the complex generalisation; the surrogate is trained on \SbS{} only.
The first-order change of the propagation constant caused by a permittivity
change $\delta\varepsilon(x,y)$ at fixed $\omega$ is~\cite{SnyderLove1983,Feynman1939}
\begin{equation}
  \delta\beta=\frac{\omega\varepsilon_0\iint\delta\varepsilon\,|\mathbf{E}|^2\,\mathrm{d}A}
                  {\iint\mathrm{Re}\left(\mathbf{E}\times\mathbf{H}^*+\mathbf{E}^*\times\mathbf{H}\right)\cdot\hat{z}\,\mathrm{d}A},
  \label{eq:pert}
\end{equation}
so a uniform index change $\delta n_i$ of region (or tensor component) $i$ gives
the confinement factor~\cite{Visser1996,Robinson2008,Sakib2019}
\begin{equation}
  \Gamma_i\equiv\left.\frac{\partial \neff}{\partial n_i}\right|_{\omega,\mathbf{L}}
  =\frac{c\,\varepsilon_0\,n_i\iint_{i}|E_i|^2\,\mathrm{d}A}{\iint\mathrm{Re}(\mathbf{E}\times\mathbf{H}^*)\cdot\hat{z}\,\mathrm{d}A}\;\ge 0 .
  \label{eq:gamma}
\end{equation}
Maxwell's equations contain no length scale: the curl--curl equation depends on
the indices only through $\omega n_i$ and on the four lengths
$\mathbf{L}=(\TLN,\Wc,\Wp,\Tp)$ only through $\omega L_k/c$~\cite{Joannopoulos2008}.
Euler's theorem then gives, at fixed material indices,
\begin{equation}
  \neff+\sum_k L_k\frac{\partial \neff}{\partial L_k} = \sum_i n_i\,\Gamma_i ,
  \label{eq:sum1}
\end{equation}
and adding the material dispersion gives $n_g=\sum_i n_{g,i}\Gamma_i$, the
equality of group and energy velocity~\cite{Biot1957}, with $n_{g,i}$ the bulk
group index of medium $i$. Subtracting the two removes the non-dispersive media
(constant-index silica and air) and leaves a first-order partial differential
equation linking the frequency derivative of $\neff$ to its gradient in the
geometry,
\begin{equation}
  n_g-\neff
  =\sum_k L_k\frac{\partial \neff}{\partial L_k}
   +\left(n_{g,s}-n_s\right)\Gamma_\mathrm{PCM}
   +\sum_{j=e,o}\left(n_{g,j}-n_j\right)\Gamma_{\mathrm{LN},j},
  \label{eq:dgi}
\end{equation}
which we call the dispersion--geometry identity. Its left side is constrained by
the spectra and its first right-hand term by the geometry sweep. The \LN{}
material terms are small ($n_{g,j}-n_j=0.045$ for $n_e$ and 0.053 for $n_o$ at
1550~nm), so an error of a few per cent in $\Gamma_{\mathrm{LN},j}$ moves the
right-hand side by less than $10^{-3}$. For the strip,
Eqs.~\eqref{eq:lorentzlorenz} and~\eqref{eq:gamma} give
\begin{equation}
  \frac{\partial \neff}{\partial f}=\Gamma_\mathrm{PCM}\,\frac{\partial n_s}{\partial f},\qquad
  \frac{\partial n_s}{\partial f}=\frac{3\left(\mathcal{L}_c-\mathcal{L}_a\right)}{2n_s\,(1-m)^2}>0,
  \label{eq:pcm}
\end{equation}
with $\mathcal{L}=(\varepsilon-1)/(\varepsilon+2)$ and
$m=f\mathcal{L}_c+(1-f)\mathcal{L}_a$, so $\neff$ rises monotonically with $f$.
Finally, perturbation theory for shifting material boundaries~\cite{Johnson2002}
gives a surface integrand
$\mathrm{d}h\,[\Delta\varepsilon_{12}|\mathbf{E}_\parallel|^2-\Delta(\varepsilon^{-1}_{12})|\mathbf{D}_\perp|^2]$
that is positive when a higher-index medium advances into a lower-index one. This
fixes the sign of $\partial\neff/\partial\Wc$, $\partial\neff/\partial\Wp$ and
$\partial\neff/\partial\Tp$ (given $\Wp\le\Wc$) but not of
$\partial\neff/\partial\TLN$, because raising the \LN{} top face puts \LN{} where
\SbS{} was beneath the strip.

The discretised solver obeys these identities closely (Table~\ref{tab:identities}),
so imposing them does not conflict with the data. $\partial\neff/\partial n_i$
was taken by central differences on a frozen grid, and
$\sum_kL_k\partial\neff/\partial L_k$ by scaling geometry, domain, grid and
absorbing layers together, which keeps the discrete problem exactly
scale-invariant. Confinement factors from the field integrals agree with the
finite-difference values to 0.1\% for $\Gamma_{\mathrm{LN},e}$ and
$\Gamma_{\mathrm{LN},o}$, and to 1.6--4.5\% for $\Gamma_\mathrm{PCM}$, whose field
integral over a thin, high-index strip depends on how the mesh resolves the field
discontinuity at its faces; the network therefore takes $\Gamma_\mathrm{PCM}$
from Eq.~\eqref{eq:pcm} applied to itself. On the grid sweep, central differences
(55 interior points) leave a \LN{} term in Eq.~\eqref{eq:dgi} of
$0.042\pm0.003$, against about 0.041 expected from
$\Gamma_{\mathrm{LN},e}\approx0.8$ and $\Gamma_{\mathrm{LN},o}\approx0.09$.

\begin{table}[!t]
  \centering
  \caption{Closure of the exact identities on the mode solver (production mesh)
    at the nominal segment ($\TLN=450$, $\Wc=1200$, $\Wp\times\Tp=600\times80$~nm,
    1550~nm), in units of index.}
  \label{tab:identities}
  \begin{tabular}{@{}p{0.52\columnwidth}cc@{}}
    \toprule
    Identity & amorphous & crystalline\\
    \midrule
    Eq.~\eqref{eq:sum1}, scale invariance & $+1.8\times10^{-5}$ & $-7\times10^{-6}$\\
    $n_g=\sum_i n_{g,i}\Gamma_i$, energy velocity & $+7.7\times10^{-5}$ & $+6.6\times10^{-5}$\\
    Eq.~\eqref{eq:dgi}, finite-difference $\Gamma_i$ & $+5.9\times10^{-5}$ & $+7.3\times10^{-5}$\\
    Eq.~\eqref{eq:dgi}, field-integral $\Gamma_i$ & $+5.9\times10^{-4}$ & $+6.6\times10^{-4}$\\
    \bottomrule
  \end{tabular}
\end{table}

\subsection{Model and training}

One multilayer perceptron maps $(\mathbf{L},f)$ to $\neff(\omega_0)$,
$n_g(\omega_0)$ and the five coefficients of $\btwo(u)$, a quartic in
$u=\omega/\omega_0-1$; $\beta$, $\beta_1$, $\neff$ and $n_g$ follow by exact
integration and $\bthree$ by exact differentiation, so $\btwo$ and $\bthree$ are
consistent by construction. The output is written as
$\mathrm{bare}(\TLN,\Wc)+\phi\,\Delta(\mathbf{L},f)$ with
$\phi=(\Tp/120~\mathrm{nm})(\Wp/900~\mathrm{nm})$, which makes the strip-free
limit exact and state-independent. A second head returns
$\Gamma_{\mathrm{LN},e}(u),\Gamma_{\mathrm{LN},o}(u)\ge0$. Input derivatives are
exact, by forward-mode automatic differentiation; because the spectra are linear
in the predicted coefficients, one Jacobian--vector product along
$(\TLN,\Wc,\Wp,\Tp,0)$ returns the whole spectrum of
$\sum_kL_k\partial\neff/\partial L_k$.

The dispersion calculus, the strip-free limit and $\Gamma\ge0$ are built into the
architecture, as in hard-constrained networks~\cite{Lu2021}. The residual of
Eq.~\eqref{eq:dgi}, with $\Gamma_\mathrm{PCM}$ from Eq.~\eqref{eq:pcm}, and the
sign conditions are penalised on unlabelled collocation points over the design
box, as in physics-informed learning~\cite{Karniadakis2021,Raissi2019,Chen2020};
fitting derivatives as well as values is a form of Sobolev
training~\cite{Czarnecki2017}. Earlier physics-aware photonic regressors embedded
the effective-index approximation~\cite{Unal2025}; every constraint used here is
exact within the assumptions above. Each residual is divided by a tolerance near
the solver noise, for example 10~\pskm{} on $\btwo$ and $10^{-3}$ on
Eq.~\eqref{eq:dgi}; the full list is in Sec.~S6.

Training uses the \SbS{} rows of the grid sweep (1546) and the space-filling rows
of the second dataset (496); the jitter set is held out entirely as truth for
Sec.~\ref{sec:dietruth}. The test split is grouped by geometry. Each member of a
five-member ensemble~\cite{Lakshminarayanan2017} is trained in three stages,
identical for all variants compared below: 20\,000 full-batch Adam
epochs~\cite{Kingma2015} on the data, 3000 epochs with the variant's own loss, and
an L-BFGS polish~\cite{Liu1989}. Baselines are a Gaussian process with a
Mat\'ern-5/2 kernel~\cite{Rasmussen2006}, gradient-boosted
trees~\cite{Friedman2001}, an unconstrained network that predicts the $\btwo$
grid and $\bthree$ as separate outputs, and the same spectral network trained
without physics terms.

\subsection{Accuracy and physical consistency}

The models are compared on interpolation inside the training domain, physical
consistency, and extrapolation outside it (Table~\ref{tab:surrcrit}). No single
model is best on all three.

\begin{figure*}[!t]
  \centering
  \includegraphics[width=0.85\textwidth]{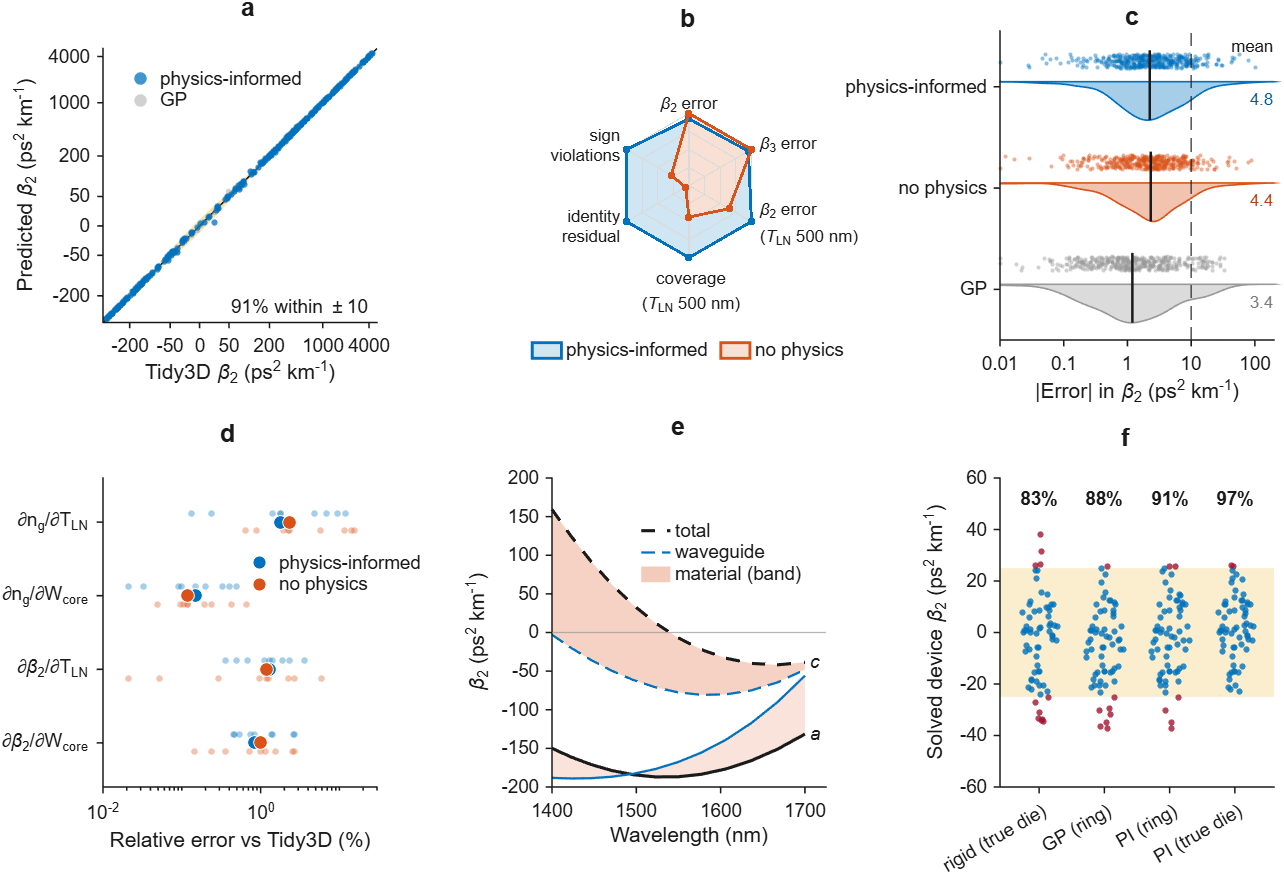}
  \caption{The physics-informed surrogate. (a)~Predicted against solved $\btwo$
    on 405 held-out rows (asinh scale); shaded: $\pm10$~\pskm. (b)~Six criteria,
    with and without physics terms, each normalised so that 1 marks the better
    model. (c)~Absolute $\btwo$ error on held-out rows. (d)~Relative error of the
    geometric derivatives against the 65 jitter dies. (e)~Waveguide and material
    parts of $\btwo$ for the 600/40~nm segment, amorphous (solid) and crystalline
    (dashed). (f)~Solved device $\btwo$ of every jitter die for four ways of
    choosing the pattern; red: outside $\pm25$~\pskm.}
  \label{fig:surr}
\end{figure*}

\paragraph{Interpolation} On the 405 held-out rows the physics-informed ensemble
predicts $\btwo$ at 1550~nm with a mean absolute error of 4.8~\pskm, with 90\% of
its predictions inside the $\pm10$~\pskm{} floor [Fig.~\ref{fig:surr}(a,c)]. The
same network without physics terms reaches 4.4~\pskm, the Gaussian process 3.4,
the unconstrained network 2.1 and gradient-boosted trees 38. All but the trees
are inside the floor, so the physics terms bring no gain in accuracy here; the
unconstrained network is the most accurate, at the price of a $\bthree$ that is
not the derivative of its own $\btwo$. A variant with only the guidance bounds
($n_g>\neff$, $n_\mathrm{SiO_2}<\neff<n_\mathrm{max}$) coincides with the network
without physics terms, because the data already satisfy the bounds.

\paragraph{Consistency} Without physics terms the network violates the sign
conditions at 6.3\% of random design points and leaves an rms residual of
$1.3\times10^{-2}$ in Eq.~\eqref{eq:dgi}, with 1.5\% of points outside its
tolerance; with them the figures are 1.1\%, $6\times10^{-4}$ and none
[Fig.~\ref{fig:surr}(b)]. These checks need the predicted $\neff$ spectrum, its
geometric gradient and the confinement factors, so they apply to the spectral
networks only. The unconstrained network's $\bthree$ differs from the derivative
of its own $\btwo$ by 0.021~\pcskm{} on average, comparable to its $\bthree$ error
of 0.018~\pcskm.

\paragraph{Extrapolation} Trained only on $\TLN\le450$~nm and asked for
$\TLN=500$~nm, the physics-informed ensemble errs by 8.3~\pskm, the network
without physics terms by 13.0 and the Gaussian process by 22.2
[Fig.~\ref{fig:surr}(b)]; the unconstrained network was not run on this split.
A separately trained spectral network without physics terms, fitted to both
claddings, gives 5.2~\pskm{} in interpolation but 36.4~\pskm{} at
$\TLN=500$~nm (Table~\ref{tab:surrcrit}). This is one held-out slice in one
direction, so the result is indicative.

\begin{table}[!t]
  \centering
  \caption{The surrogates on the three criteria. Interpolation: mean absolute
    error (MAE, \pskm) of $\btwo$ at 1550~nm on the 405 held-out rows.
    Consistency: mean gap between the predicted $\bthree$ and the derivative of
    the predicted $\btwo$ (\pcskm), and, at random design points, the fraction of
    sign conditions violated and the rms residual of Eq.~\eqref{eq:dgi}.
    Extrapolation: MAE at $\TLN=500$~nm after training on $\TLN\le450$~nm.
    PI, physics-informed; GP, Gaussian process; MLP, unconstrained network;
    n/a, not defined for the model; --, not computed. $^a$Network predicting
    the $\btwo$ spectrum without physics terms, trained on both claddings with a
    separate grouped split (606 held-out rows; 756 rows at $\TLN=500$~nm). Its
    early stopping used validation rows that included $\TLN=500$~nm, so its
    extrapolation error is likely an underestimate.}
  \label{tab:surrcrit}
  \setlength{\tabcolsep}{2.8pt}
  \begin{tabular}{@{}lccccc@{}}
    \toprule
     & Interp. & \multicolumn{3}{c}{Consistency} & Extrap.\\
    \cmidrule(lr){3-5}
    Model & MAE & $\bthree$ gap & Sign (\%) & Residual & MAE\\
    \midrule
    PI         & 4.8 & 0     & 1.1 & $6\times10^{-4}$   & 8.3\\
    No physics & 4.4 & 0     & 6.3 & $1.3\times10^{-2}$ & 13.0\\
    GP         & 3.4 & --    & n/a & n/a                & 22.2\\
    MLP        & 2.1 & 0.021 & n/a & n/a                & --\\
    Trees      & 38  & --    & n/a & n/a                & --\\
    ANN (non-physics)$^a$ & 5.2 & n/a & n/a & n/a       & 36.4\\
    \bottomrule
  \end{tabular}
\end{table}

\paragraph{Derivatives and calibration} Against the derivatives fitted to the 65
jitter dies, the median relative error of $\partial n_g/\partial\TLN$ is 2.0\%
with the physics terms and 2.3\% without, that of $\partial\btwo/\partial\TLN$
1.2\% for both, and both width derivatives are reproduced to about 1\%
[Fig.~\ref{fig:surr}(d)]. The raw ensemble spread is narrower than the error: its
90\% interval covers 69\% of the held-out values (normalised errors with a
standard deviation of 2.9), against 92\% for the Gaussian process, and at
$\TLN=500$~nm 76\%, against 64\% without the physics terms. The pattern choice
below therefore uses the spread inflated by a factor of 1.93, calibrated for 90\%
coverage on the held-out rows.

Differentiating Eq.~\eqref{eq:dgi} in $\omega$ splits $\btwo$ exactly into a
waveguide part
$c^{-1}\mathrm{d}(\neff+\sum_kL_k\partial\neff/\partial L_k)/\mathrm{d}\omega$
and a material part, which the surrogate evaluates from its own gradient
[Fig.~\ref{fig:surr}(e)]. For the 600/40~nm segment at 1550~nm the amorphous
$\btwo$ of $-187$~\pskm{} (Tidy3D: $-187.3$) is $-168$ waveguide and $-19$
material. On crystallisation both parts rise by 90~\pskm, so half of the
switching contrast is the dispersion of crystalline \SbS{} itself and half is
the change of the mode.

\section{Die-Specific Inference and Trimming}
\label{sec:die}

\subsection{Selecting the segment set}
\label{sec:yield}

The four of the eleven segment types placed on the mask are chosen once, before
fabrication. All 330 four-element combinations were ranked by the fraction of
Monte Carlo dies (below) brought inside specification. The best set,
$\Wp/\Tp=300/80$, 600/40, 900/40 and 900/80~nm, is tied by the second-best and
followed closely by the next, so several sets perform similarly (Sec.~S4 and
Fig.~S7 of the \supp{}).

\subsection{A linear Monte Carlo}

Each of 5000 synthetic dies per material draws a ridge-width error and a
film-thickness error from normal distributions with $\sigma_W=14.9$~nm and
$\sigma_T=8.0$~nm, representing electron-beam linewidth control and commercial
wafer thickness uniformity. The as-fabricated $\btwo$ is the nominal value plus a
first-order expansion in the two errors with the sensitivities of
Fig.~\ref{fig:sens}, and the trimmer picks the bit pattern that brings $\btwo$
closest to zero. A die for which no pattern lands inside $\pm25$~\pskm{} is
untrimmable.

\begin{figure*}[!t]
  \centering
  \includegraphics[width=0.85\textwidth]{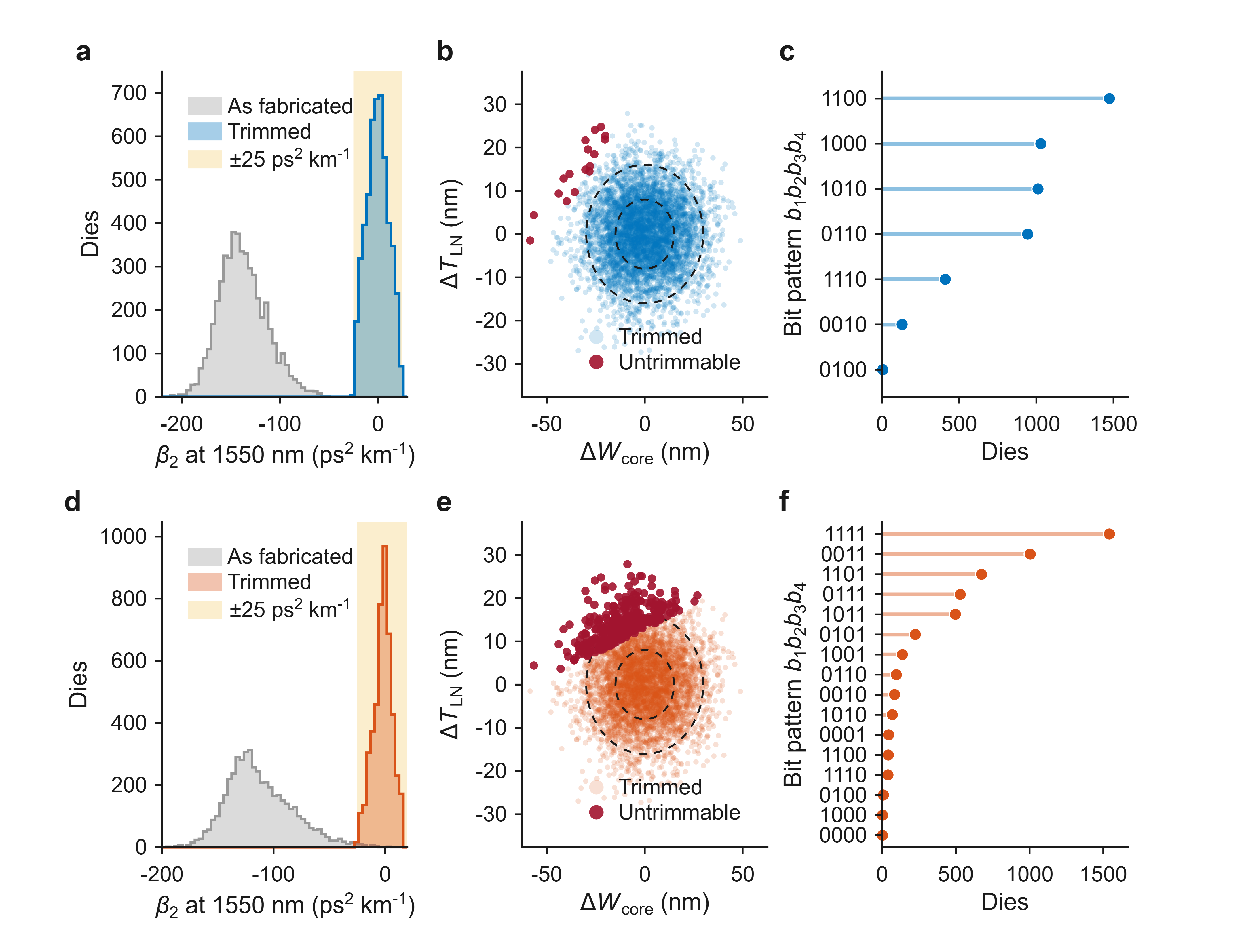}
  \caption{Monte Carlo over 5000 dies per material; top \SbS{}, bottom \SbSe{}.
    (a,d)~$\btwo$ before (grey) and after (colour) trimming; shaded:
    $\pm25$~\pskm. (b,e)~Error plane with $1\sigma$ and $2\sigma$ process
    contours; untrimmable dies in dark red. (c,f)~Dies per bit pattern.}
  \label{fig:mc}
\end{figure*}

For \SbS{}, 99.66\% of dies are trimmed into specification and the standard
deviation of $\btwo$ falls from 24.2 to 11.1~\pskm, at a median added loss of
$1.2\times10^{-7}$~\dbcm{} (Fig.~\ref{fig:mc}). The residual spread is set by the
discreteness of the 16 states (Sec.~\ref{sec:libdesign}). The nominal design is
biased anomalous, at $\btwo\approx-137$~\pskm, because crystallisation moves
$\btwo$ only in the normal direction at these strip thicknesses; the trim removes
this designed offset and the process spread together. The untrimmable 0.34\% of
\SbS{} dies occupy one corner of the $(\Delta\Wc,\Delta\TLN)$ plane
[Fig.~\ref{fig:mc}(b)], a region that could be screened at wafer test. Seven of
the 16 patterns are used. \SbSe{} uses all 16 and leaves 6.7\% of dies
untrimmable, at a median loss of 3.1~\dbcm. The neglected quadratic terms have a
median of 0.4 to 0.6~\pskm{} and reach 11.3~\pskm{} at three standard deviations
(Sec.~S5), so the linear model holds for the bare ridge over most of the
distribution; whether the segment states follow it is tested next.

\subsection{A full-wave check at fabricated geometries}
\label{sec:dietruth}

\begin{figure*}[!t]
  \centering
  \includegraphics[width=0.8\textwidth]{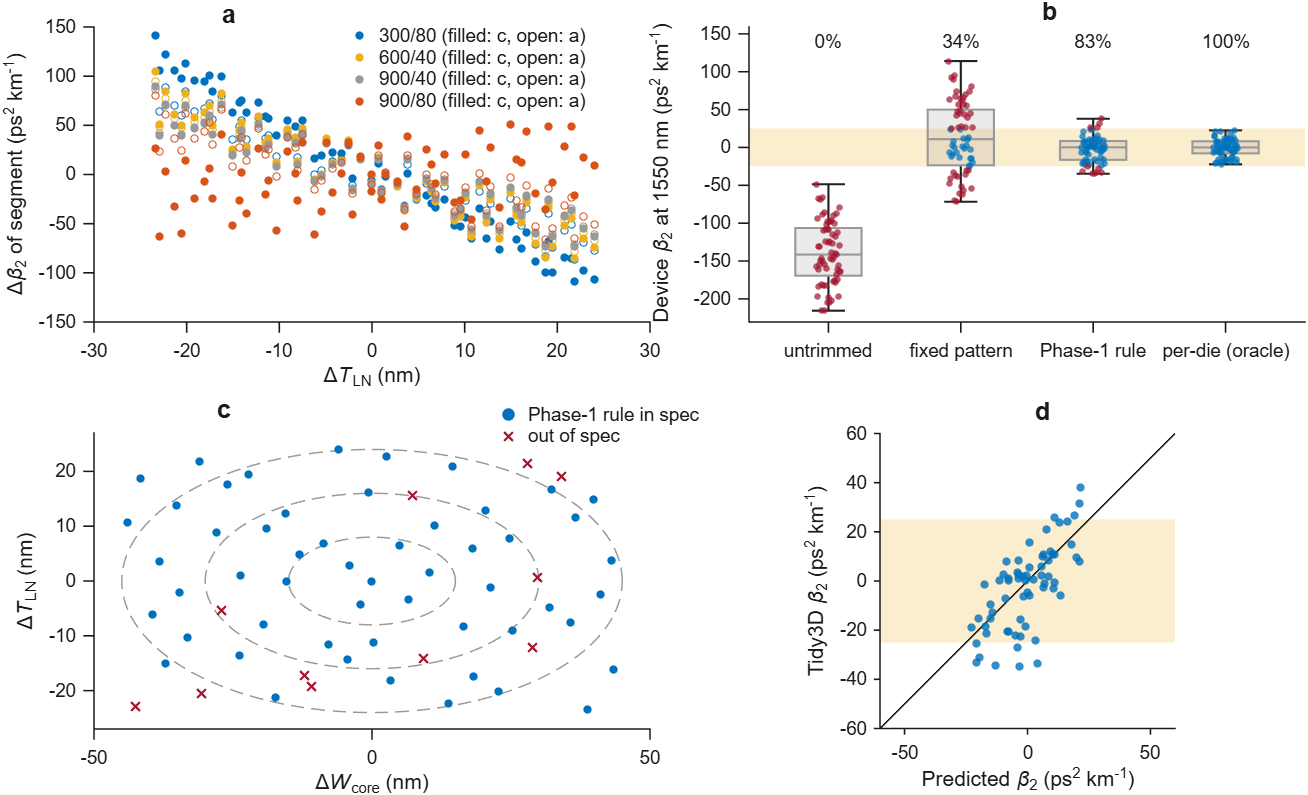}
  \caption{Mode solves at 65 die geometries in the $\pm3\sigma$ box. (a)~Change of
    each segment state's $\btwo$ from the nominal die against the film-thickness
    error; filled: crystalline. (b)~Device $\btwo$ for four ways of choosing the
    pattern; red: out of specification. (c)~Error plane for the rigid-shift rule
    with 1, 2 and $3\sigma$ contours. (d)~Device $\btwo$ predicted by the
    rigid-shift rule against the solved value.}
  \label{fig:dietruth}
\end{figure*}

The Monte Carlo treats a die error as a rigid shift: a die is assumed to move
every pattern of the nominal library by as much as it moves the bare ridge. Each
of the 65 jitter dies has its own solved library, so this can be tested directly.
The segment states do not move together [Fig.~\ref{fig:dietruth}(a)]. The
amorphous segments follow the bare ridge, with $\partial\btwo/\partial\TLN$
between $-2.1$ and $-3.4$~\pskm{} per nanometre against $-3.7$ for the bare
ridge, but the crystalline 900/80~nm segment has a slope of $+0.7$~\pskm{} per
nanometre and a width sensitivity of 1.1~\pskm{} per nanometre, against 0.4 to
0.7 for the others, and its switching contrast ranges from 1937 to 2096~\pskm{}
over the box.

Four ways of choosing the pattern were scored on the solved device $\btwo$ of each
die, Eq.~\eqref{eq:betamix} with equal lengths [Fig.~\ref{fig:dietruth}(b) and
Table~\ref{tab:dietruth}]. With no pattern written no die is in specification, as
designed, and the single pattern best on the nominal die, $1100$, helps only dies
near the nominal. The rigid-shift rule, given each die's error exactly, brings
83\% of the box into specification (94\% weighted by the process density, with an
effective sample size of 22 dies). It mispredicts the device $\btwo$ by a
standard deviation of 12.2~\pskm{} and by up to 38~\pskm{}
[Fig.~\ref{fig:dietruth}(d)], and all eleven of its failures lie more than
$1.8\sigma$ from the nominal die [Fig.~\ref{fig:dietruth}(c)]. Choosing from the
die's own library brings all 65 dies in, with a residual standard deviation of
11.8~\pskm.

\begin{table}[!t]
  \centering
  \caption{Device $\btwo$ at 1550~nm on the 65 solved jitter dies for four ways of
    choosing the bit pattern. In spec: $|\btwo|\le25$~\pskm; weighted by the
    process density.}
  \label{tab:dietruth}
  \begin{tabular}{@{}p{0.36\columnwidth}cccc@{}}
    \toprule
    Pattern choice & \multicolumn{2}{c}{In spec (\%)} & \multicolumn{2}{c}{SD of $\btwo$ (\pskm)}\\
     & box & weighted & box & weighted\\
    \midrule
    None (0000) & 0 & 0 & 41.2 & 22.6\\
    Fixed, nominal best (1100) & 34 & 63 & 46.7 & 25.7\\
    Rigid shift of nominal library & 83 & 94 & 17.1 & 13.7\\
    Library of the die (oracle) & 100 & 100 & 11.8 & 11.4\\
    \bottomrule
  \end{tabular}
\end{table}

\subsection{Inferring the die error and choosing the pattern}
\label{sec:infer}

The die error is read from test rings whose free spectral range gives the group
index at 1500, 1550 and 1600~nm, with a $1\sigma$ resolution of $2\times10^{-4}$.
Each jitter die's true $n_g$ plus measurement noise is inverted on the surrogate,
the pattern with the highest probability of landing within $\pm25$~\pskm{} is
chosen, and the outcome is scored on the Tidy3D device $\btwo$ of that die
[Fig.~\ref{fig:surr}(f)]. With one bare ring the surrogate brings 91\% of the
dies (59 of 65; 88\% weighted by the process density) into specification in a
single noise draw, against 88\% (87\% weighted) when the Gaussian process chooses
from the same inferred geometry. Given the true geometry the surrogate brings
every die in, against 83\% (94\% weighted) for the rigid-shift rule with the same
information, so the per-die model removes the error of the rigid shift and what
remains is the inference. One ring recovers the geometry with rms errors of
4.8~nm in $\TLN$ and 10.0~nm in $\Wc$, and three of the six failures miss the band
by less than 1~\pskm.

\subsection{A second test ring}
\label{sec:rings}

One bare ring constrains mainly one combination of the two errors: a linearised
Fisher analysis with the process prior gives posterior standard deviations of
3.3~nm in $\TLN$ and 6.0~nm in $\Wc$ with a correlation of 0.99, because the
group index moves with $\TLN$ and $\Wc$ in nearly the same ratio at all three
wavelengths. A second bare ring 300~nm wider helps only partly (1.3 and 2.7~nm,
correlation 0.97). A ring carrying a phase-change strip needs no new structure,
since the mask already holds eight strip states, and its group index at every
jitter die is in the jitter solves, so the loop is scored entirely on Tidy3D. A
crystalline strip draws the mode towards the cladding and changes the ratio of
the two sensitivities. Table~\ref{tab:rings} lists every pairing, with the
surrogate's own group-index error ($2.9\times10^{-4}$ rms on the held-out rows)
added to the ring resolution as noise. Every crystalline strip ring removes most
of the correlation, and the crystalline 900/80~nm ring gives the smallest error,
0.41~nm in $\TLN$ and 0.91~nm in $\Wc$ [Fig.~\ref{fig:rings}(a)].

\begin{table}[!t]
  \centering
  \caption{Die error from the test rings at the nominal die, and the loop scored
    on the Tidy3D device $\btwo$ of the 65 jitter dies (ten noise draws each).
    In spec: $|\btwo|\le25$~\pskm; a, amorphous; c, crystalline strip.}
  \label{tab:rings}
  \begin{tabular}{lccccc}
    \toprule
    Rings & sd($\TLN$) & sd($\Wc$) & Corr. & In spec & Weighted\\
          & (nm) & (nm) & & (\%) & (\%)\\
    \midrule
    bare            & 4.4  & 8.1  & 0.98    & 88.8 & 93.0\\
    bare + 300/80 a & 1.17 & 1.37 & 0.75    & 88.8 & 88.1\\
    bare + 300/80 c & 0.66 & 0.81 & 0.01    & 92.9 & 91.8\\
    bare + 600/40 a & 1.08 & 1.27 & 0.68    & 90.9 & 90.8\\
    bare + 600/40 c & 0.66 & 0.85 & 0.04    & 94.5 & 92.8\\
    bare + 900/40 a & 0.96 & 1.17 & 0.57    & 92.9 & 90.6\\
    bare + 900/40 c & 0.58 & 0.90 & $-0.07$ & 96.0 & 96.4\\
    bare + 900/80 a & 0.52 & 0.89 & $-0.23$ & 96.2 & 97.0\\
    bare + 900/80 c & 0.41 & 0.91 & $-0.53$ & 96.8 & 99.1\\
    true geometry   & --   & --   & --      & 100  & 100\\
    \bottomrule
  \end{tabular}
\end{table}

In the loop each ring is read at the three wavelengths, the die error is taken
from the posterior on a grid over $\pm4\sigma$, and the pattern is chosen from 48
posterior samples; each die is measured ten times with independent noise. One
bare ring brings 88.8\% of the trials within $\pm25$~\pskm{} (93.0\% weighted).
Adding the crystalline 900/80~nm ring raises this to 96.8\% (99.1\% weighted),
with 3.2\% false successes, and the rms errors of the recovered geometry fall from
6.1 and 11.7~nm to 1.7 and 4.2~nm [Fig.~\ref{fig:rings}(b)]. The pair is
insensitive to the ring resolution: at $1\times10^{-4}$ and $5\times10^{-4}$ it
gives 96.9\% and 96.6\%, while one bare ring falls from 91.5\% to 79.2\%. The
remaining failures therefore appear to be set mainly by the surrogate's own
group-index error rather than by the measurement.

\subsection{Sensitivity to a systematic solver offset}
\label{sec:offset}

The success rates above are scored on the solver's own $\btwo$ and so inherit the
numerical floor of Sec.~\ref{sec:conv}. A systematic error shifts every die the
same way. To bound its effect, a common offset $\delta$ was added to the solved
device $\btwo$ of every die after the pattern had been chosen, modelling a
fabricated device that differs from the solver by $\delta$ while the surrogate,
trained on the solver, does not (Table~\ref{tab:offset}). With the two rings the
success rate is 96.8\% at $\delta=0$, 99.8\% at $+5$ and 98.6\% at
$+10$~\pskm, and 86.0\% at $-5$ and 73.5\% at $-10$~\pskm. The asymmetry follows
from the trimmed distribution, with mean $-4.8$~\pskm{} and standard deviation
11.6~\pskm: a positive offset recentres it, and a negative one pushes its lower
tail out of the band. Tightening the band has the same effect: the two rings bring
86.0\% of the trials within $\pm20$~\pskm{} and 72.2\% within $\pm15$~\pskm,
against 78.8\% and 60.6\% for one bare ring. An offset could be measured on a
first fabricated die and removed by shifting the target of the pattern choice.

\begin{table}[!t]
  \centering
  \caption{Success rate within $\pm25$~\pskm{} (\%) on the 65 jitter dies, ten
    noise draws each, with a common offset $\delta$ added to the solved device
    $\btwo$. w, weighted by the process density.}
  \label{tab:offset}
  \begin{tabular}{@{}lccccc@{}}
    \toprule
    $\delta$ (\pskm) & $-10$ & $-5$ & 0 & $+5$ & $+10$\\
    \midrule
    Bare + 900/80~c & 73.5 & 86.0 & 96.8 & 99.8 & 98.6\\
    \quad w & 73.4 & 87.6 & 99.1 & 99.9 & 98.7\\
    Bare ring only & 70.0 & 84.8 & 88.8 & 87.7 & 84.9\\
    \quad w & 73.1 & 88.1 & 93.0 & 90.9 & 91.1\\
    \bottomrule
  \end{tabular}
\end{table}

\subsection{Designing the segment library with exact derivatives}
\label{sec:libdesign}

Because the surrogate is differentiable in the strip geometry, the segment library
itself can be designed. A library is four strips $(\Wp,\Tp)$ inside the training
box and four length fractions of the 2~mm device. It is scored on 1024 dies drawn
from the process distribution, with the pattern chosen by the calibrated ensemble
and each ensemble member taken in turn as the truth. A smooth version of this
yield is maximised with Adam (four starts of 150 steps per loss weight) at
$\pm10$~\pskm, where designs differ; at $\pm25$~\pskm{} every library considered,
including the current mask, brings more than 99.5\% of the dies in. Junction loss
enters through a proxy, $k\sum_j(\Delta n_{\mathrm{eff},j})^2$ over the junctions
of a pattern, with $k=22.1$~dB fitted to the single-segment eigenmode-expansion
runs of Sec.~\ref{sec:eme}; it reproduces those runs (0.06 to 2.45~dB) and the
four-segment runs (0.45 to 1.72~dB) to within 0.19~dB. On the current mask the
yield estimate matches the Tidy3D dies at $\pm25$~\pskm{} (100\% for both) and
lies 3 to 5 points above them otherwise (58.5\% against 53.8\% at
$\pm10$~\pskm, and 99.7\% against 96.8\% with the two rings).

\begin{figure*}[!t]
  \centering
  \includegraphics[width=0.9\textwidth]{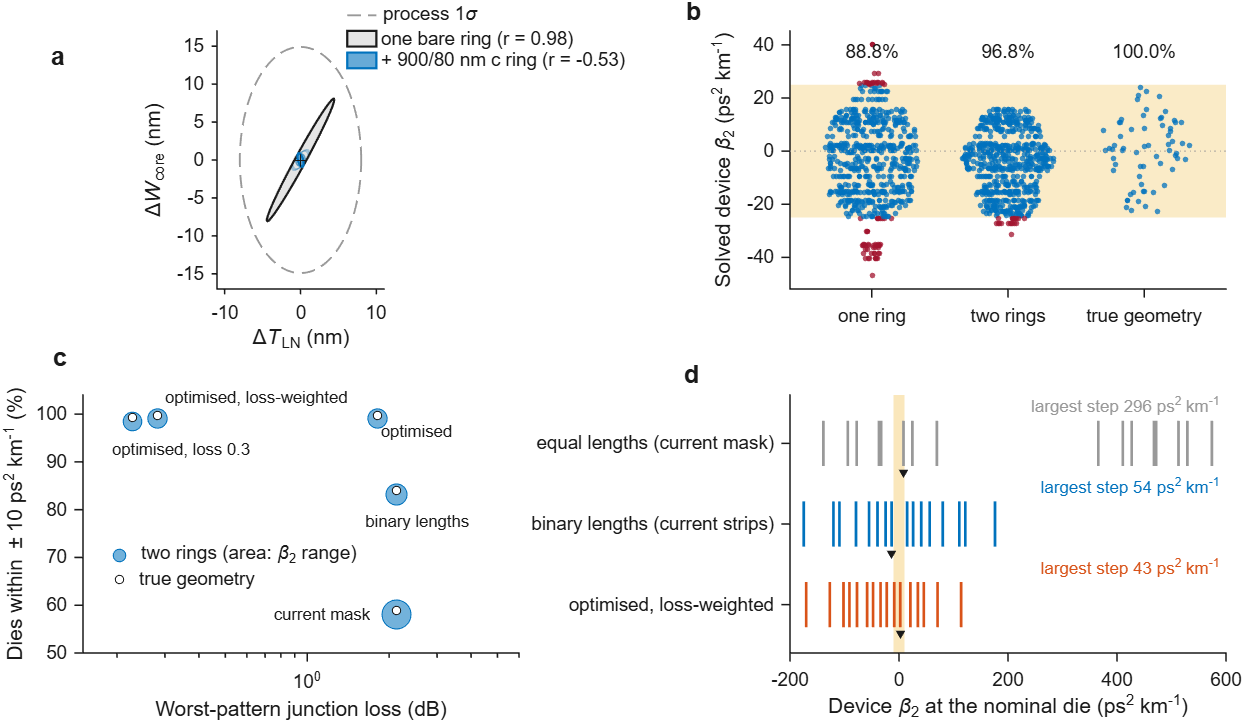}
  \caption{Test rings and library design. (a)~Posterior $1\sigma$ ellipses of the
    die error: process prior (dashed), one bare ring, and bare plus the
    crystalline 900/80~nm ring (thin: the other strip rings). (b)~Solved device
    $\btwo$ of the jitter dies after the pattern choice with one ring, two rings
    and the true geometry; shaded: $\pm25$~\pskm. (c)~Model yield at
    $\pm10$~\pskm{} against worst-pattern junction loss; bubble area: $\btwo$
    range. (d)~The 16 levels at the nominal die for equal, binary and optimised
    lengths; shaded: $\pm10$~\pskm.}
  \label{fig:rings}
\end{figure*}

At $\pm10$~\pskm{} the current mask brings 58.9\% of the dies in with the true
geometry (58.1\% with two rings), because its 16 levels cluster
[Fig.~\ref{fig:rings}(d)]. Keeping its strips but making the lengths binary
(0.13, 0.53, 0.27 and 0.07 of the device for the 300/80, 600/40, 900/40 and
900/80~nm strips) spreads the levels and gives 84.0\% (83.1\%). Optimising strips
and lengths together gives 99.8\% (99.1\%). With the junction loss weighted at
0.1 per dB, the optimiser replaces the 900/80~nm strip, whose crystalline step
dominates the loss, by a 448/64~nm strip; the library 300/74, 543/40, 899/42 and
448/64~nm with lengths 0.13, 0.50, 0.23 and 0.14 keeps 99.8\% (99.0\%) while the
worst pattern loses 0.28~dB instead of 2.13~dB [Fig.~\ref{fig:rings}(c)]. Its
patterns span $-169$ to 116~\pskm, narrower than the $-137$ to 576~\pskm{} of the
current mask but more than 100~\pskm{} on either side of zero. At 1.5 times the
process spread it keeps 97.1\% (96.0\%), against 59.7\% for the current mask. In
the model, an offset of up to the 10~\pskm{} floor would keep dies placed within
$\pm10$~\pskm{} inside $\pm20$~\pskm. The optimised strips lie inside the
training box; a mode solve of the new strips and an eigenmode-expansion run with
the new lengths are the next checks.

\subsection{Summary of success rates}
\label{sec:summary}

Table~\ref{tab:success} collects the success rates, which differ in what they are
scored on. The linear Monte Carlo is scored on its own model; on full-wave solves
the same rigid-shift rule reaches 94\% of the weighted dies rather than 99.66\%.
The jitter box over-represents the tails of the process distribution (effective
sample size 22 dies), which is why the weighted values are higher. The
library-design row is scored on the surrogate ensemble, at a tighter band, and
awaits a full-wave check.

\begin{table}[!t]
  \centering
  \caption{Success rates of the pattern-choice rules. Band $\pm25$~\pskm{} except
    where stated; w, weighted by the process density; LM, linear model; T3D,
    Tidy3D device $\btwo$ of the jitter dies; ens., surrogate ensemble as truth.
    Trials: dies, times ten noise draws for the ring rows and five ensemble
    members for the last row.}
  \label{tab:success}
  \begin{tabular}{@{}p{0.40\columnwidth}cccc@{}}
    \toprule
    Pattern choice & Truth & Trials & In spec & w\\
     & & & (\%) & (\%)\\
    \midrule
    Rigid shift, exact error & LM & 5000 & 99.66 & --\\
    Rigid shift, exact error & T3D & 65 & 83 & 94\\
    Library of the die (oracle) & T3D & 65 & 100 & 100\\
    Surrogate, exact geometry & T3D & 65 & 100 & 100\\
    Surrogate, bare ring & T3D & 650 & 88.8 & 93.0\\
    Surrogate, bare + 900/80~c & T3D & 650 & 96.8 & 99.1\\
    \quad with $\delta=-10$~\pskm & T3D & 650 & 73.5 & 73.4\\
    \quad with $\delta=+10$~\pskm & T3D & 650 & 98.6 & 98.7\\
    Optimised library, two rings, $\pm10$~\pskm & ens. & 1024 & 99.0 & --\\
    \bottomrule
  \end{tabular}
\end{table}

\section{Thermal Feasibility of Laser Writing}
\label{sec:laser}

The trim is applied once, after the die has been measured, so it needs no
integrated heater: a segment could be crystallised by a focused beam, as done for
\SbS{} films with a 532~nm laser, where continuous-wave exposure recrystallised
the film within 7 to 600~ms at 1 to 30~mW~\cite{Laprais2024}. Only the
amorphous-to-crystalline direction is needed, so no melt-quench step is involved.
The estimate below is thermal only: it determines the absorbed fraction, whether
the strip can be held above its crystallisation temperature without melting, and
whether the neighbouring strip stays below its own. Crystallisation kinetics and
the uniformity and repeatability of a scanned write are not modelled.

\begin{figure*}[!t]
  \centering
  \includegraphics[width=0.8\textwidth]{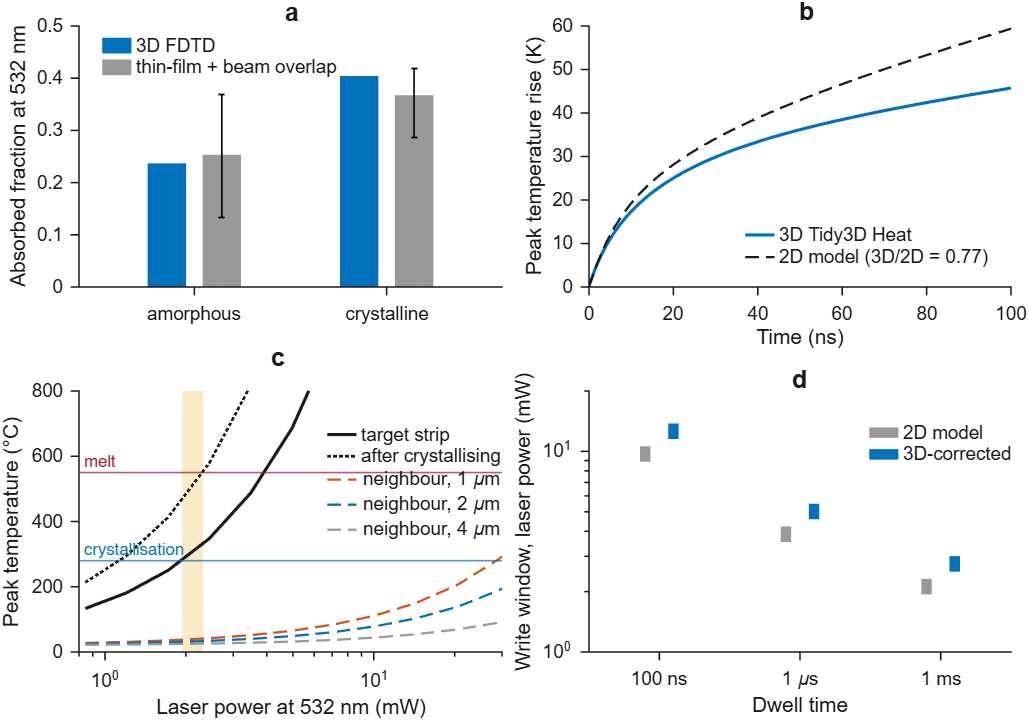}
  \caption{Thermal model of a 532~nm write. (a)~Absorbed fraction in
    the target strip, FDTD and thin-film estimate. (b)~Peak temperature rise for
    3~mW absorbed, three- and two-dimensional models. (c)~Peak temperature against
    incident power for a 1~ms dwell: target strip amorphous (solid) and
    crystallised (dotted), neighbour at 1, 2 and 4~\um{} (dashed); shaded:
    predicted write window. (d)~Predicted window against dwell time.}
  \label{fig:laser}
\end{figure*}

\paragraph{Absorption and heating}
A three-dimensional FDTD solve at 532~nm with a Gaussian beam of 0.6~\um{} waist
centred on one 600/80~nm strip gives an absorbed fraction of 0.237 with the strip
amorphous and 0.404 crystalline, and $2\times10^{-6}$ in the neighbouring strip
1.5~\um{} away [Fig.~\ref{fig:laser}(a)]. Transient heat conduction was solved
in the plane through the two strip axes; a three-dimensional transient solve for
one case gives a peak rise 0.77 times that of the slice
[Fig.~\ref{fig:laser}(b)], and powers are quoted both from the slice and divided
by that ratio. The model, its constants and a thin-film check of the absorption
are given in Sec.~S7 and Table~S4.

\paragraph{A predicted write window}
With constant material properties the temperature rise is linear in absorbed
power, and three conditions fix the window: the amorphous strip must reach the
crystallisation temperature, 280~$^\circ$C~\cite{Laprais2024}; the strip must stay
below the melting point, 550~$^\circ$C~\cite{CRC2016}, after it has crystallised
and absorbs $A_\mathrm{c}/A_\mathrm{a}=1.71$ times more of the beam; and the
neighbour must stay at least 50~K below the crystallisation temperature. The
second condition binds and limits the window to a factor
$(T_\mathrm{m}-T_0)/[(T_\mathrm{X}-T_0)A_\mathrm{c}/A_\mathrm{a}]=1.20$ in power,
independent of dwell time, so the writing power would have to be held to about
$\pm9\%$. For a 1~ms dwell, the regime of Ref.~\cite{Laprais2024}, the predicted
window is 1.9 to 2.3~mW of incident power from the slice and an estimated 2.5 to
3.0~mW after the three-dimensional correction [Fig.~\ref{fig:laser}(c,d)], which
rests on a single three-dimensional case. At the top of the window the
neighbouring strip heats by 21, 13 and 6~K for gaps of 1, 2 and 4~\um. Shorter
dwells need 4.6 to 5.5~mW at 1~\us{} and 11.5 to 13.8~mW at 100~ns after
correction; there the crystallisation kinetics of the film~\cite{HoffAPR2025}
also enter. The window assumes constant material properties and complete
crystallisation once the threshold is reached; both, and the effect of the
writing conditions on the crystalline index, need measurement.

\section{Discussion}
\label{sec:discussion}

The largest uncertainty in the two-ring success rate is the numerical floor. If
the solver error is systematic, the rate lies between 73.5\% and 99.8\% for
offsets within $\pm10$~\pskm{} (Table~\ref{tab:offset}); a calibration on a first
fabricated die, or a library designed for a tighter band
(Sec.~\ref{sec:libdesign}), would recover most of the margin. The \SbS{} optical
constants also vary between laboratories~\cite{GutierrezIScience2022}, which
shifts the absolute dispersion values, and the negligible material loss holds for
the adopted constants ($k=0$) rather than for any deposited film. In the present
model the dominant optical cost is the insertion loss of the abrupt junctions,
mostly at the crystalline 900/80~nm segment; a converged figure needs a larger
mode basis or a finite-difference time-domain calculation.

The second test ring is the main practical element of the inference. Film and
width errors are nearly degenerate in the group index of a bare ring, and a ring
carrying a crystalline strip, already on the mask, removes most of the
degeneracy. With two rings the remaining failures follow the surrogate's
group-index error, so more training data near the nominal die is the most direct
way to reduce them. Inside the training domain the physics terms do not improve
accuracy; they add consistency between $\btwo$ and $\bthree$, derivatives with the
correct signs, and a smaller error in the one extrapolation tested. The
identities are exact for the lossless, non-magnetic guide solved here and would
need their complex form for an absorbing cladding.

The sweep is of a fully etched ridge, whereas most thin-film lithium niobate
processes leave a slab; a partial etch changes the confinement and therefore
$\btwo$, and the numbers here apply to the geometry as swept. The write step is
modelled only through absorption and heat flow. The time at temperature needed
for full crystallisation, the uniformity along a 500~\um{} segment written by a
scanned spot, the repeatability from segment to segment and any change of the
crystalline index with the writing conditions are informed by work on \SbS{}
kinetics~\cite{HoffAPR2025}, laser writing~\cite{Laprais2024} and
back-end-of-line integration~\cite{WeiNatCommun2024}, but are not predicted here.
Because the trim is one-way, cycling endurance does not enter. All results are
numerical.

\section{Conclusion}
\label{sec:conclusion}

In simulation, a binary \SbS{} cladding divided into a few independently switched
segments moves the group-velocity dispersion of a thin-film lithium niobate
waveguide over a range large compared with the error of a commercial wafer, with
two degrees of freedom, so $\btwo$ and $\bthree$ need not be traded against each
other. Four segments reach $-140$ to 580~\pskm{} in $\btwo$, and eigenmode
expansion indicates that the junctions leave the length-weighted model intact to
within 1.24~\pskm.

Because the segment states move with the die error, the bit pattern is best
chosen from a model of the die. A spectral surrogate constrained by exact
identities of the lossless Maxwell problem provides that model with consistent
$\btwo$ and $\bthree$, geometric derivatives within about 2\% of the solver, and
an error of 8.3 rather than 13.0~\pskm{} at an unseen film thickness.
Read through a bare ring and a ring carrying the crystalline 900/80~nm strip, it
brings 96.8\% of the dies (99.1\% weighted) within $\pm25$~\pskm{} on full-wave
truth, and every die when the geometry is known; the solver floor, rather than
the inference, sets the main uncertainty.

The same derivatives make the segment library a design variable: jointly
optimised strips and lengths bring 99\% of the model dies within $\pm10$~\pskm{}
while the worst-case junction loss falls from 2.1 to 0.3~dB. A thermal model
predicts a candidate write window of 2.5 to 3.0~mW at 532~nm for a 1~ms dwell;
the kinetics, uniformity and repeatability of the write remain to be measured. A
fabricated trimmer with its two test rings is the next step, and the same
procedure could address other dispersion targets, such as a comb pump or a
frequency-conversion bandwidth.

\section*{Author Contributions}

\textbf{S.D.}: conceptualization, methodology, software, formal analysis,
writing of the original draft.
\textbf{A.K.M.}: machine-learning modelling, writing (review and editing),
manuscript preparation.
\textbf{S.S.}: supervision.

\section*{Disclosures}
The authors declare no conflicts of interest.

\section*{Data Availability}
The simulation results and the scripts that produce every figure and table are available from the authors on reasonable request.

\section*{Acknowledgments}

\bibliographystyle{IEEEtran}
\bibliography{references}

\end{document}


\maketitle

\noindent This document contains Sections S1--S7, Figures S1--S7 and Tables
S1--S4. Equation, figure and section numbers without the prefix S refer to the
main text.


\section{Material models and numerical settings}

\subsection{Optical constants}
\begin{figure}[H]
  \centering
  \includegraphics[width=0.8\textwidth]{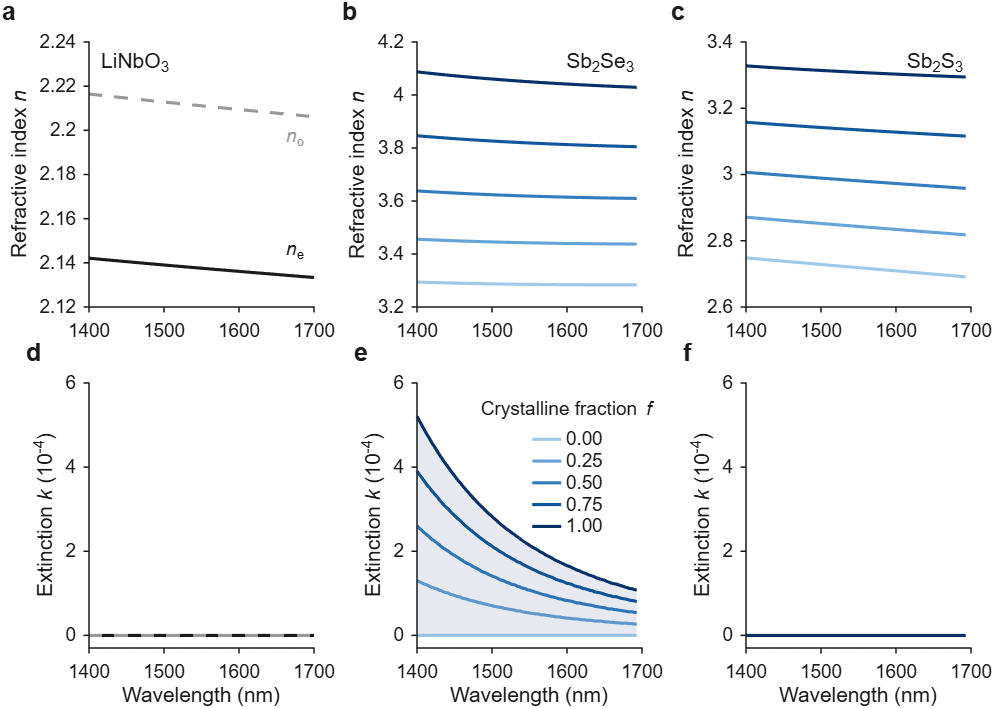}
  \caption{Bulk optical constants used by the mode solver: refractive index
    (top) and extinction coefficient (bottom). (a,d)~\LN{} extraordinary (solid)
    and ordinary (dashed), Sellmeier fit of Ref.~\cite{ZelmonJOSAB1997}.
    (b,e)~\SbSe{} and (c,f)~\SbS{} from Ref.~\cite{DelaneyAFM2020} at five
    crystalline fractions $f$, intermediate states by the Lorentz--Lorenz rule.}
  \label{fig:matdisp}
\end{figure}

The \LN{} film uses the three-oscillator Sellmeier fit of Zelmon \textit{et
al.}~\cite{ZelmonJOSAB1997} for congruent undoped \LN{} (0.4 to 5.0~\um,
21~$^\circ$C), with the extraordinary index for the quasi-TE mode of the X-cut
film. The phase-change layers use the ellipsometry $n$ and $k$ of Delaney
\textit{et al.}~\cite{DelaneyAFM2020} for the two end states, tabulated from 194
to 1693~nm and combined by the Lorentz--Lorenz rule, Eq.~(1) of the main text, at
intermediate crystalline fractions $f$ (Fig.~\ref{fig:matdisp}). Above 1693~nm
the tables are extrapolated, over an interval in which $n$ changes by less than
$10^{-4}$ per nanometre. A two-phase Bruggeman medium built from the same end
states differs from the Lorentz--Lorenz rule by up to 0.045 in $n$ for \SbSe{}
and 0.03 for \SbS{}.

\subsection{Mesh and mode profiles}
\begin{figure}[H]
  \centering
  \includegraphics[width=\textwidth]{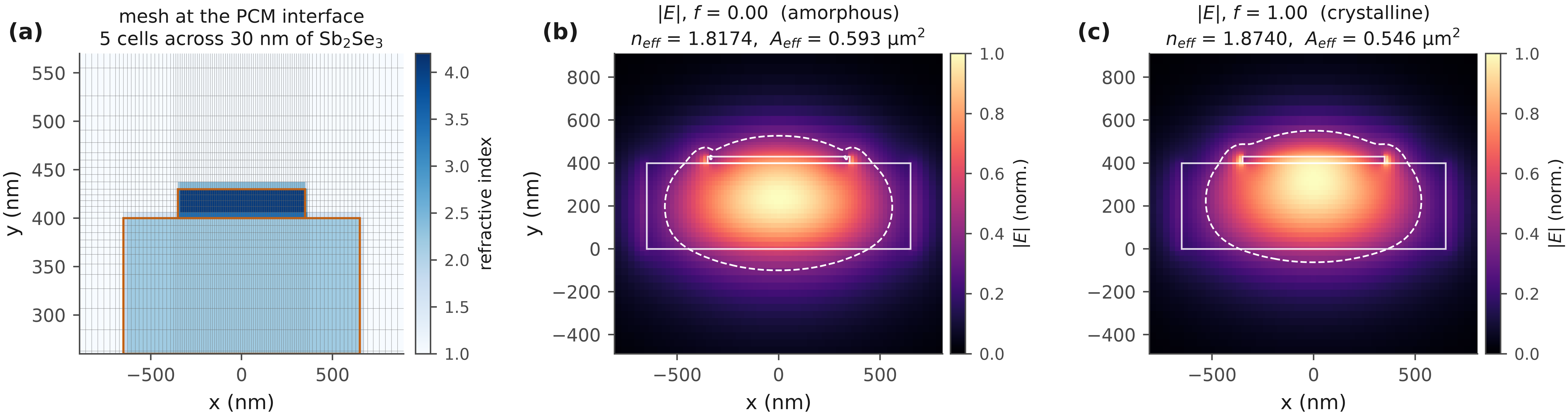}
  \caption{Mesh and mode. (a)~Index distribution at the phase-change interface,
    showing the cells that resolve the strip. (b,c)~Normalised $|E|$ of the
    quasi-TE mode with the strip amorphous and crystalline.}
  \label{fig:modes}
\end{figure}

The cross-sectional domain is 6.0 $\times$ 4.0~\um{} at 32 grid steps per
wavelength, with snapping points at the silica, \LN{}, phase-change and air
interfaces so that every layer boundary coincides with a cell edge
[Fig.~\ref{fig:modes}(a)]. The band edges of the 1400 to 1700~nm reporting band
are interpolated rather than extrapolated, because the solves extend 20~nm beyond
each end. The selected polynomial order and the fit residual are recorded for
every row.

\subsection{Pole-residue fit tolerance}
Each dispersive medium is fitted to a pole-residue model before the solve, and
the fit tolerance had a larger effect than any meshing parameter tested. In a diagnostic run
with an RMS tolerance of $2\times10^{-4}$, the bulk \LN{} $\btwo$ at 1550~nm came
out at 373~\pskm; tightening it to $10^{-7}$ gave 99.5~\pskm, against
99.2~\pskm{} from a cubic spline through the raw data (Table~\ref{tab:matfit}).
Every production solve in both datasets uses the $10^{-7}$ tolerance with at most
four poles, and every fitted medium was checked against a spline of its own
table; the loose tolerance appears only in this diagnostic. For \SbS{}, $k$ is
zero in the adopted data, so the computed \SbS{} losses are the numerical floor
of the pole fit.

\begin{table}[H]
  \centering
  \caption{Bulk \LN{} $\btwo$ at 1550~nm against the pole-residue fit tolerance.
    Production solves use $10^{-7}$; the first row is a diagnostic.}
  \label{tab:matfit}
  \begin{tabular}{@{}p{0.55\columnwidth}c@{}}
    \toprule
    Treatment of the index table & $\btwo$ at 1550~nm (\pskm)\\
    \midrule
    Pole-residue fit, RMS tolerance $2\times10^{-4}$ & 373\\
    Pole-residue fit, RMS tolerance $10^{-7}$ & 99.5\\
    Cubic spline through the raw table & 99.2\\
    \bottomrule
  \end{tabular}
\end{table}

\subsection{Convergence}
Convergence was assessed at a geometry on a zero crossing, where a relative
criterion is meaningless, so absolute agreement in \pskm{} was used
(Table~\ref{tab:conv}). Enlarging the domain moves $\btwo$ by 0.5~\pskm{} while
shrinking it moves $\btwo$ by 41.9~\pskm, which fixes the production domain at
6.0 $\times$ 4.0~\um; lowering the resolution to 25 steps per wavelength moves it
by 56.1~\pskm. The settings used are the smallest that are converged, and the
residual uncertainty on $\btwo$ near a zero crossing is about $\pm10$~\pskm.

Raising the frequency count from 11 to 21 changed $\btwo$ by at most 8~\pskm{}
across the band. A fixed second-order fit deviated by up to 1280~\pskm{} at the
band edges, and a fixed fourth-order fit, accurate at 1550~nm, by about
117~\pskm{} there, which is why the fit order is adaptive.

\begin{table}[H]
  \centering
  \caption{Convergence of $\btwo$ at a zero crossing (\pskm). Baseline: 6.0
    $\times$ 4.0~\um{} domain, 32 steps per wavelength, 11 frequencies, adaptive
    fit order. Lower block: largest deviation across the band.}
  \label{tab:conv}
  \begin{tabular}{@{}p{0.40\columnwidth}
                  >{\centering\arraybackslash}p{0.22\columnwidth}
                  >{\centering\arraybackslash}p{0.28\columnwidth}@{}}
    \toprule
    Variation & $\btwo$ at 1550~nm & Change from baseline\\
    \midrule
    Baseline & $+3.9$ & \\
    Domain 5.0 $\times$ 3.5~\um & $-38.0$ & 41.9\\
    Domain 7.0 $\times$ 5.0~\um & $+3.4$ & 0.5\\
    Resolution 25 steps per wavelength & $-52.2$ & 56.1\\
    \midrule
    Variation & & Largest deviation across band\\
    \midrule
    21 frequencies rather than 11 & & up to 8\\
    Fixed second-order fit & & up to 1280\\
    Fixed fourth-order fit & & about 117\\
    \bottomrule
  \end{tabular}
\end{table}

\subsection{Quality filter and coverage}
A solved geometry enters the dataset only if a guided quasi-TE mode exists, if
$\neff$ falls monotonically with wavelength across the extended band, and if the
$\neff$ fit residual is at most $10^{-6}$. The monotonicity test catches mode
anticrossing: when the crystalline strip is thick enough to guide a mode of its
own, the tracked mode exchanges character with it and $\neff(\omega)$ acquires
structure that a low-order polynomial should not be differentiated across.
Rejected nodes are recorded with the criterion that rejected them.

The sweep produced 3012 converged solutions, 1546 for \SbS{} and 1466 for
\SbSe{}, covering 92.0\% and 87.3\% of their grids (Table~\ref{tab:excluded}). For
\SbS{}, 65 of the 134 rejected nodes lie at the thickest strip, $\Tp=120$~nm.
Restricting to $\Tp\le80$~nm, the range the trimmer uses, the rejection rate
falls from 8.0\% to 5.5\%, and within the design region ($\Tp\le80$~nm,
$\TLN\ge400$~nm) to 4.6\%. No reported result rests on an excluded node: the
representative design point of the main text is present at all five fractions,
the four mask segment types are present in both states, and the segment library
uses solved geometries only.

\begin{table}[H]
  \centering
  \caption{Coverage of the two sweep grids. A node is one
    $(\TLN,\Wc,\Wp,\Tp,f)$ combination.}
  \label{tab:excluded}
  \begin{tabular}{lcccc}
    \toprule
    Material & Grid nodes & Converged & Rejected & Coverage\\
    \midrule
    \SbS{}  & 1680 & 1546 & 134 & 92.0\%\\
    \SbSe{} & 1680 & 1466 & 214 & 87.3\%\\
    Both    & 3360 & 3012 & 348 & 89.6\%\\
    \bottomrule
  \end{tabular}
\end{table}

\section{The second, full-wave dataset}
\begin{figure}[H]
  \centering
  \includegraphics[width=0.9\textwidth]{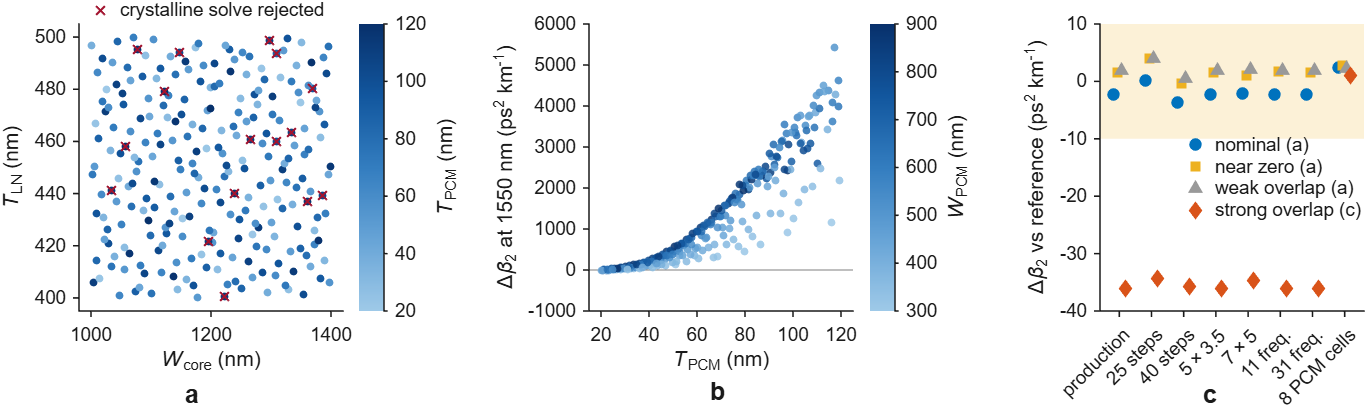}
  \caption{The full-wave dataset used by the surrogate. (a)~The 256 space-filling
    geometries, coloured by $\Tp$; crosses mark the 16 failed crystalline solves.
    (b)~Switching contrast at 1550~nm against $\Tp$, coloured by $\Wp$.
    (c)~Change of $\btwo$ from the finest setting for each one-at-a-time
    variation, at four test geometries; shaded: $\pm10$~\pskm.}
  \label{fig:p2data}
\end{figure}

Figure~\ref{fig:p2data}(a) shows the 256 space-filling geometries. Of their 512
solves, 16 failed the quality filter: 15 crystalline anticrossings and one fit
residual above $10^{-6}$. The switching contrast grows with $\Tp$ as in the grid
sweep, from a median of 13~\pskm{} at $\Tp\le30$~nm to 2630~\pskm{} above 80~nm
[Fig.~\ref{fig:p2data}(b)]. The median $\Gamma_\mathrm{PCM}$ at 1550~nm is 0.07
with the strip amorphous and 0.13 crystalline.

Convergence was re-checked against the finest setting tried (48 steps per
wavelength, a 7 $\times$ 5~\um{} domain and 31 frequencies) at geometries chosen
to stress the solver [Fig.~\ref{fig:p2data}(c)]. With the strip amorphous, every
one-at-a-time variation stays within 4~\pskm{} of the reference. The crystalline
strong-overlap geometry ($\Wp/\Tp=900/80$~nm, $\btwo\approx2000$~\pskm) is off by
$-36$~\pskm{} until the number of cells across the strip is doubled, which brings
it to $+1$~\pskm, so the error there is set by the vertical resolution of the
strip. These checks set the numerical floor used in the main text, the
larger of $\pm10$~\pskm{} and 2\% of $|\btwo|$.

\section{Dispersion maps of the as-deposited waveguide}
\begin{figure}[H]
  \centering
  \includegraphics[width=0.7\columnwidth]{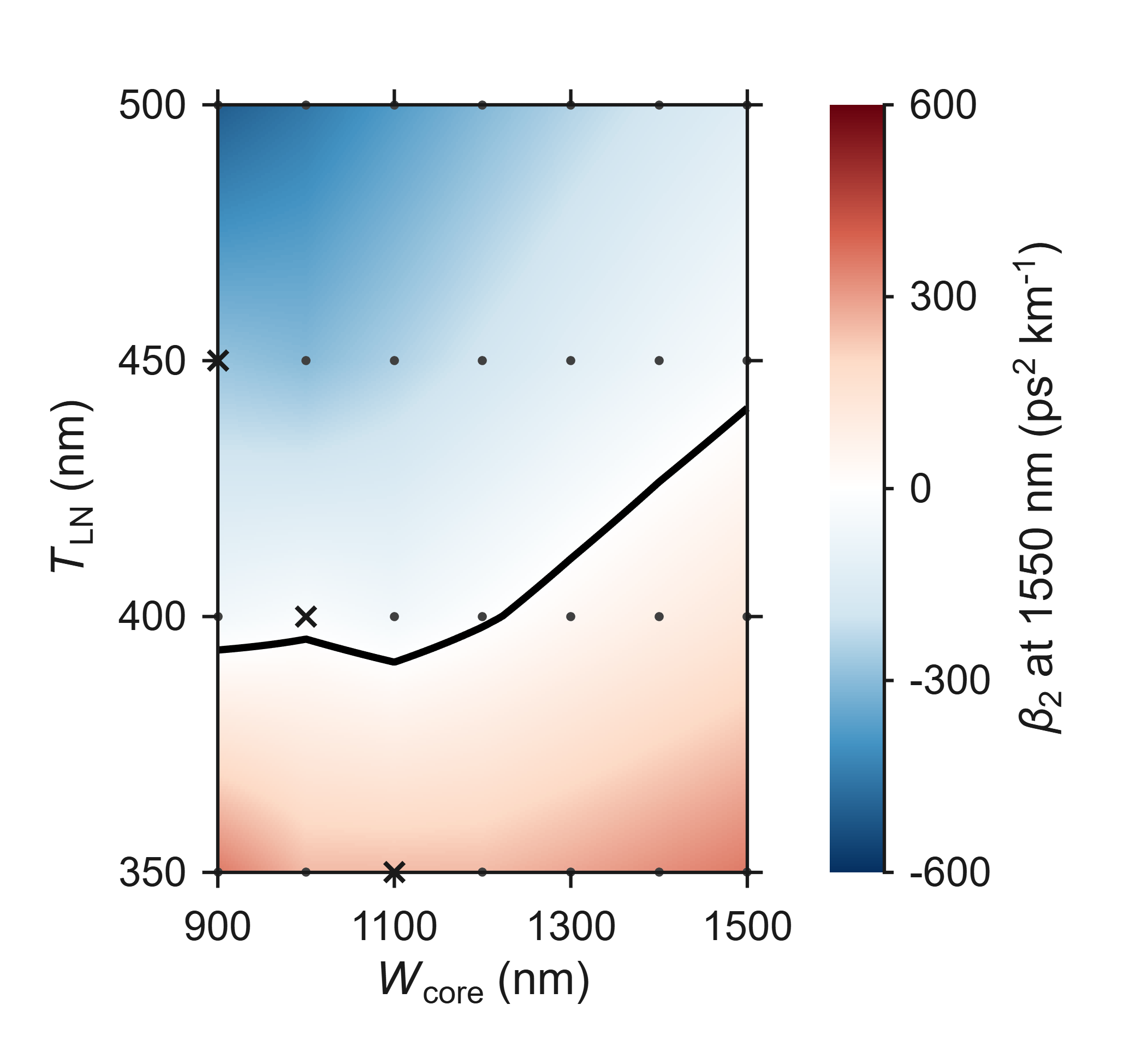}
  \caption{$\btwo$ at 1550~nm of the as-deposited \SbS{}-clad waveguide over
    $(\Wc,\TLN)$ at $\Wp=600$~nm, $\Tp=40$~nm, interpolated between solved
    geometries (dots); crosses are nodes rejected by the quality filter. The solid
    contour is $\btwo=0$.}
  \label{fig:zdwmap}
\end{figure}

\begin{figure}[H]
  \centering
  \includegraphics[width=0.9\textwidth]{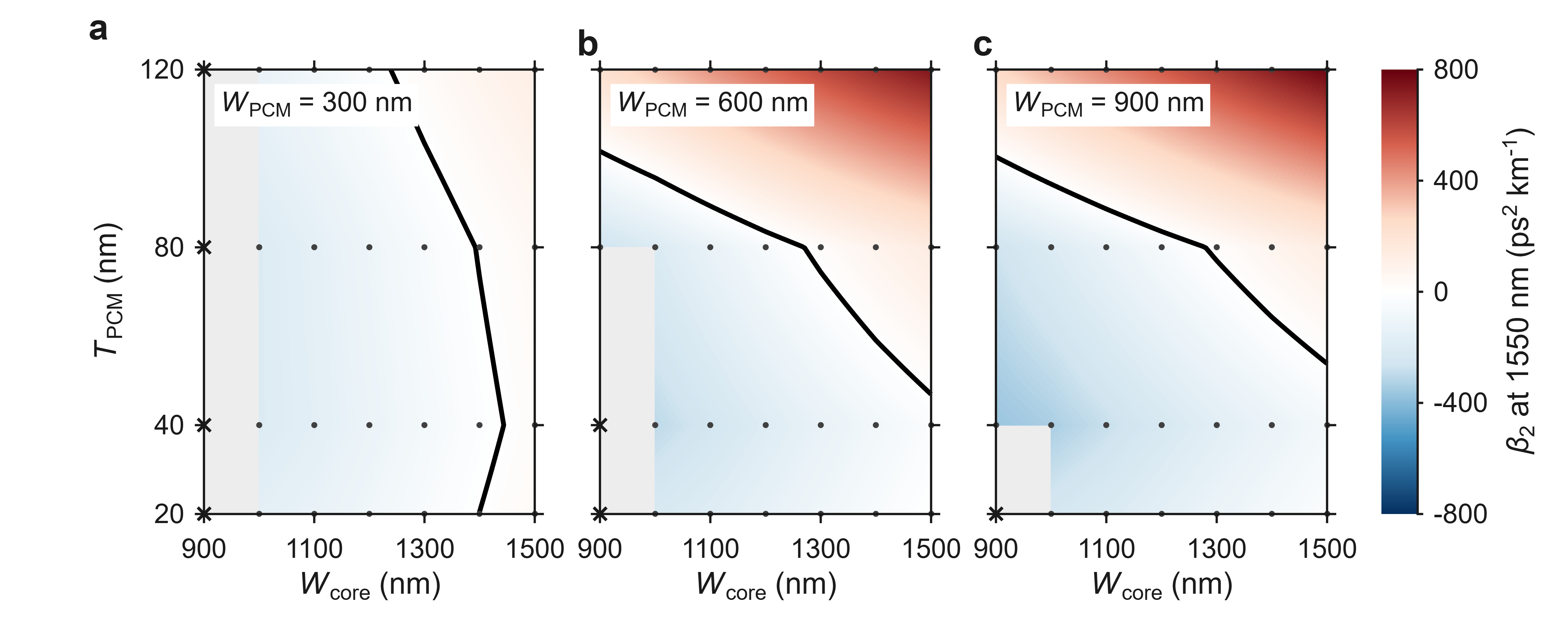}
  \caption{Amorphous $\btwo$ at 1550~nm over $(\Wc,\Tp)$ at $\TLN=450$~nm for
    (a)~$\Wp=300$, (b)~600 and (c)~900~nm. Black contours: $\btwo=0$. Dots are
    solved geometries, crosses rejected nodes; grey regions have no solved
    neighbours.}
  \label{fig:designmaps}
\end{figure}

\begin{figure}[H]
  \centering
  \includegraphics[width=0.8\textwidth]{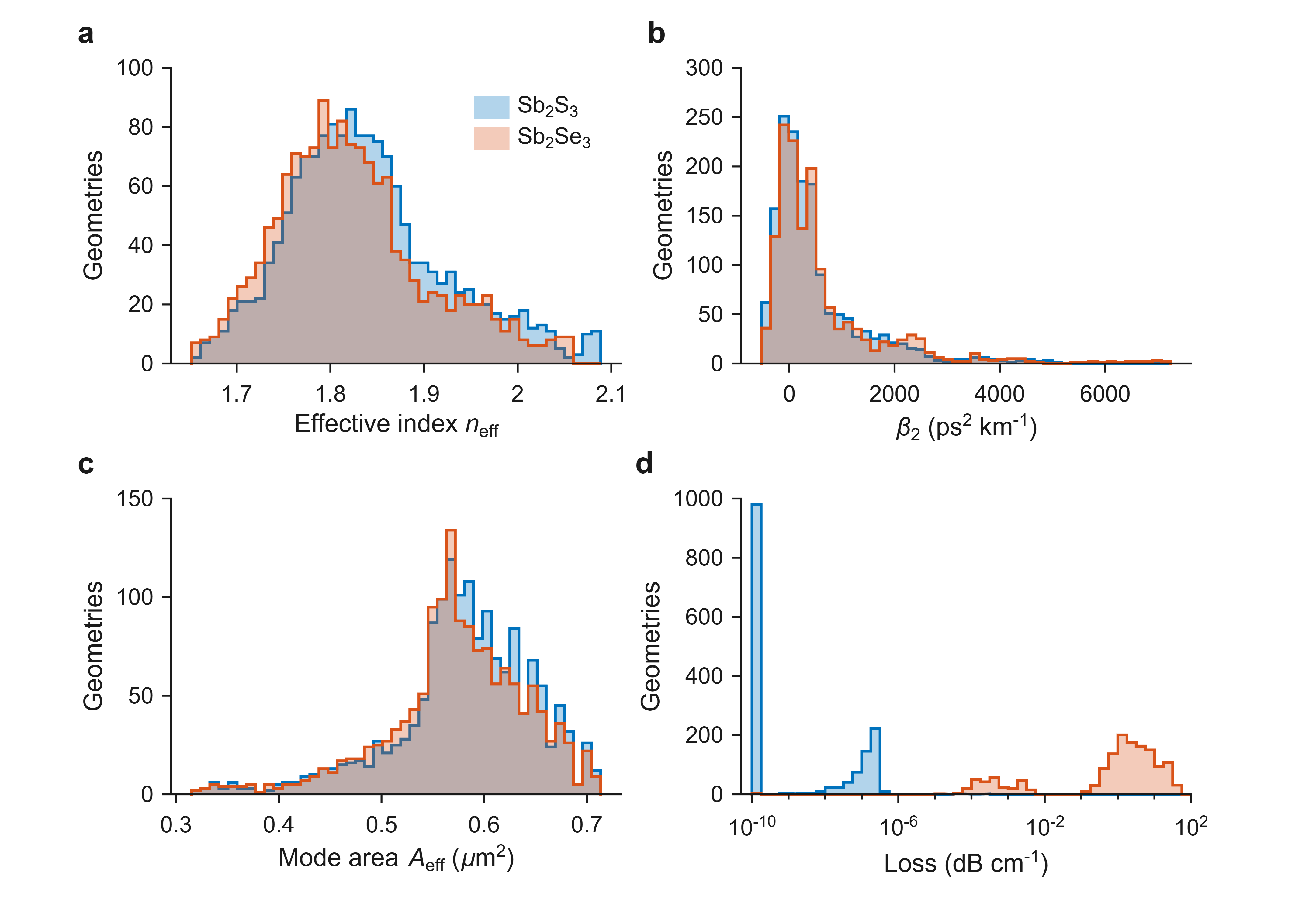}
  \caption{Distributions over the converged sweep at 1550~nm, all crystalline
    fractions. (a)~Effective index. (b)~$\btwo$; positive is normal.
    (c)~Effective mode area. (d)~Propagation loss, clamped at the
    $10^{-10}$~\dbcm{} floor of the pole fit.}
  \label{fig:overview}
\end{figure}

Figure~\ref{fig:zdwmap} shows the zero-dispersion locus over the ridge
dimensions, Fig.~\ref{fig:designmaps} its dependence on the strip, and
Fig.~\ref{fig:overview} the distributions of the main modal quantities over the
converged sweep.

\section{Selecting the segment set}
\begin{figure}[H]
  \centering
  \includegraphics[width=0.85\textwidth]{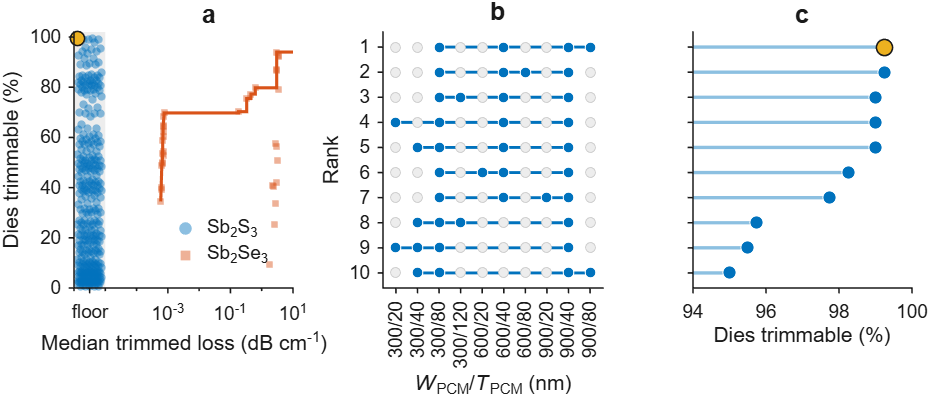}
  \caption{Selecting the four segment types on the mask. (a)~Each candidate set:
    fraction of Monte Carlo dies trimmed into specification against median loss;
    \SbS{} sets lie at the numerical floor (shaded), the selected set is ringed,
    and the step line is the \SbSe{} Pareto front. (b)~The ten best \SbS{} sets:
    segment types used (left) and fraction of dies trimmed (right).}
  \label{fig:segdesign}
\end{figure}

Which four of the eleven segment types go on the mask is decided once, before
fabrication. All 330 four-element combinations were evaluated against the linear Monte
Carlo of the main text and ranked by the fraction of dies brought inside specification, with
median added loss as a tie-break. The best set, $\Wp/\Tp=300/80$, 600/40, 900/40
and 900/80~nm, brings 99.25\% of dies into specification; the second-best ties
it, the seventh reaches 97.75\% and the tenth 95.0\%, so several sets perform
similarly (Fig.~\ref{fig:segdesign}). For \SbS{} all losses are below the
numerical floor and
the selection is made on yield; for \SbSe{} it trades several per cent of yield
against several \dbcm.

\section{Validity of the linearisation}
The expansion uses finite-difference sensitivities taken across the sweep grid
(50~nm in thickness, 100~nm in width), over process spreads of 8.0 and 14.9~nm.
Second differences along the thickness axis give
$|\partial^2\btwo/\partial\TLN^2|$ with a median of 0.025~\pskm{} per nm$^2$
(90th percentile 0.040); the width curvature is two orders of magnitude smaller.
Propagated through the same normal draw, the neglected quadratic term has a median
of 0.4 to 0.6~\pskm, well inside the $\pm10$~\pskm{} numerical floor, and reaches
3.0 to 4.9~\pskm{} at the 95th percentile and 11.3~\pskm{} at three standard
deviations. The linear model is therefore accurate across the bulk of the
distribution for the bare ridge; the full-wave check at fabricated geometries in
the main text tests whether the segment states follow it.

\section{Surrogate training and evaluation}

\subsection{Training tolerances}
Each residual in the training loss is divided by its tolerance: $3\times10^{-4}$ on absolute $\neff$, about twice the
geometry-correlated noise of the automatic mesh, $2\times10^{-4}$ on $n_g$,
10~\pskm{} on $\btwo$ and $10^{-3}$ on Eq.~(7) of the main text. The rows of the second
dataset carry the field-integral confinement factors, fitted with a tolerance of
$5\times10^{-3}$ for $\Gamma_{\mathrm{LN},e}$ and $\Gamma_{\mathrm{LN},o}$ and of
5\% (plus $5\times10^{-3}$) for the $\Gamma_\mathrm{PCM}$ implied by
Eq.~(8) of the main text. Where they are not labelled, the cladding share
$n_\mathrm{SiO_2}\Gamma_\mathrm{SiO_2}+\Gamma_\mathrm{air}$ implied by
Eq.~(6) of the main text is kept between 0.03 and 0.35 and
$\Gamma_{\mathrm{LN},o}/\Gamma_{\mathrm{LN},e}$ between 0.05 and 0.25, windows
around the finite-difference values at the nominal segment.

\subsection{Evaluation metrics}
The quantities in Table~3 of the main text are computed as follows.

\paragraph{Interpolation} The test split holds 20\% of the rows, grouped by
geometry so that no geometry appears in both the training and the test set
(405 rows). The interpolation error is the mean absolute difference between the
ensemble-mean prediction and the solved $\btwo$ at 1550~nm.

\paragraph{$\bthree$ gap} The predicted $\btwo$ spectrum is differentiated
numerically in $\omega$ on the wavelength grid of the $\btwo$ output, and the gap
is the mean of $|\bthree-\mathrm{d}\btwo/\mathrm{d}\omega|$ at 1550~nm over the
held-out rows, with both terms taken from the same model. In the spectral
networks $\bthree$ is the analytic derivative of $\btwo$, so their gap reflects
only the finite-difference error and is below $10^{-3}$~\pcskm.

\paragraph{Sign conditions} The consistency checks use 2048 points drawn uniformly
over the design box, with $\Wp\le\Wc$, $f$ between 0 and 1 and about 10\% of the
points without a strip; no solver data exist at these points. Forward-mode
automatic differentiation gives $\partial\neff/\partial f$ and
$L_k\,\partial\neff/\partial L_k$ for $\Wc$, $\Wp$ and $\Tp$ at every wavelength of
the spectral grid. $\TLN$ is excluded because the sign of its derivative is not
fixed. A point and wavelength pair counts as a violation if any of the four
derivatives is below $-10^{-4}$, and the reported value is the fraction of such
pairs, averaged over the five ensemble members.

\paragraph{Identity residual} On the same points, the residual of Eq.~(7) of the
main text,
\begin{equation}
  r = n_g-\neff-\sum_k L_k\frac{\partial\neff}{\partial L_k}
      -\left(n_{g,s}-n_s\right)\Gamma_\mathrm{PCM}
      -\sum_{j=e,o}\left(n_{g,j}-n_j\right)\Gamma_{\mathrm{LN},j},
\end{equation}
is evaluated with $\Gamma_\mathrm{PCM}$ from Eq.~(8) of the main text and
$\Gamma_{\mathrm{LN},j}$ from the confinement-factor head of the network. The
reported value is the rms of $r$ over points and wavelengths, averaged over the
ensemble members. The fraction of points outside tolerance quoted in the main
text counts the points at which the \LN{} term, $r$ before its last sum, lies
outside the range allowed by the material dispersion of \LN{} by more than the
$10^{-3}$ tolerance. Both checks need the $\neff$ spectrum, its geometric
gradient and the confinement factors, so they are defined only for the spectral
networks.

\paragraph{Extrapolation} Both spectral ensembles are retrained from the start on
the rows with $\TLN\le450$~nm, with the same three-stage schedule, and the
Gaussian process is refitted on the same rows with the same kernel. All three
are scored by the mean absolute error of $\btwo$ at 1550~nm on the
$\TLN=500$~nm rows. The unconstrained network and the boosted trees were not
retrained.

\paragraph{Coverage} The 90\% coverage is the fraction of rows whose error lies
within $\pm1.645$ times the predicted standard deviation.

\section{Thermal model of the laser write}
\paragraph{Absorption}
A three-dimensional FDTD solve at 532~nm with a Gaussian beam of 0.6~\um{} waist
centred on one 600/80~nm strip gives an absorbed fraction of 0.237 with the strip
amorphous and 0.404 crystalline, and $2\times10^{-6}$ in the neighbouring strip
1.5~\um{} away [main-text Fig.~12(a)]. A thin-film calculation of the same
stack, averaged over the buried-oxide fringes and multiplied by the beam overlap,
gives 0.25 and 0.37. The indices at 532~nm, $3.50+0.21i$ amorphous and
$4.18+0.75i$ crystalline, are close to the ellipsometry of
Ref.~\cite{Laprais2024} ($3.55+0.18i$ and $4.72+0.74i$).

\paragraph{Heating}
Transient heat conduction was solved in the plane through the two strip axes with
a finite-volume model and implicit time stepping, with the absorbed power
deposited in the target strip over a Gaussian footprint of 1.2~\um{} $1/e^2$
radius along its axis. The constants are listed in Table~\ref{tab:thermal}. A
two-dimensional slice has no lateral heat spreading; a three-dimensional transient
solve in Tidy3D for one case (3~mW absorbed uniformly in a 10~\um{} strip for
100~ns) gives a peak rise of 45.7~K against 59.4~K from the slice, a ratio of 0.77
[main-text Fig.~12(b)]. Powers are quoted both from the slice and divided by
that ratio.

\begin{table}[H]
  \centering
  \caption{Thermal constants of the write-window model: conductivity $\kappa$ and
    volumetric heat capacity $\rho c_p$.}
  \label{tab:thermal}
  \begin{tabular}{@{}lccl@{}}
    \toprule
    Material & $\kappa$ (W\,m$^{-1}$\,K$^{-1}$) & $\rho c_p$ (MJ\,m$^{-3}$\,K$^{-1}$) & Source\\
    \midrule
    \SbS{} & 1.16 & 1.62 & \cite{Laprais2024}\\
    \LN{} & 4.6 & 2.88 & nominal; see~\cite{Bashir2024}\\
    SiO$_2$ & 1.38 & 1.64 & \cite{CRC2016}\\
    Si & 148 & 1.63 & \cite{Glassbrenner1964,CRC2016}\\
    \midrule
    \multicolumn{4}{@{}l}{\SbS{}: crystallisation 280~$^\circ$C~\cite{Laprais2024}, melting 550~$^\circ$C~\cite{CRC2016}}\\
    \bottomrule
  \end{tabular}
\end{table}

\bibliographystyle{unsrt}
\bibliography{references}